\documentclass[preprint,aps,showpacs,showkeys]{revtex4}
\usepackage{epsfig,amsmath,amssymb}
\begin{document}

\title{Dynamic properties of quantum spin chains:
       Simple route to complex behavior}
\author{Taras Verkholyak$^{1}$,
        Oleg Derzhko$^{1}$,
        Taras Krokhmalskii$^{1}$,
        and
        Joachim Stolze$^{2}$}
\affiliation{$^1$Institute for Condensed Matter Physics,
             National Academy of Sciences of Ukraine,
             1 Svientsitskii Street, L'viv-11, 79011, Ukraine\\
             $^2$Institut f\"{u}r Physik, Universit\"{a}t Dortmund,
             44221, Dortmund, Germany}

\date{\today}

\begin{abstract}
We examine dynamic structure factors of spin-1/2 chains
with nearest-neighbor interactions of $XX$ and Dzyaloshinskii-Moriya type,
and with periodic and random changes in the sign of these interactions.
This special kind of inhomogeneity can be eliminated from the Hamiltonian
by suitable transformation of the spin variables.
As a result, the dynamic structure factors of periodic or random chains
can be computed from those of the uniform chains. Using the exact analytical
and precise numerical results available for the uniform systems we illustrate
the effects of regular alternation or random disorder
on dynamic structure factors of quantum spin chains.
\end{abstract}



\maketitle


\section{Introduction. Jordan-Wigner fermions and dynamic quantities}
\label{sec1}
\setcounter{equation}{0}

Quantum spin chains have received much attention during the
last more than 70 years for several reasons.
First,
they provide an excellent ground for studying rigorously quantum many-particle phenomena.
Second,
owing to the tremendous progress in material sciences
(as well as
the recent availability of optical lattices for trapping atoms in artificial crystals)
many real-life systems, which can be modeled as quantum spin chains invented by theoreticians,
have become available.
That opens the possibility to compare the results of accurate theoretical calculations with experimental data.
Dynamic quantities for quantum spin chains are of special interest and importance.
On the one hand,
their study,
as a rule,
is a harder problem in comparison with similar studies of static quantities.
On the other hand,
dynamic quantities are related to experimental data obtained in scattering and resonance experiments
which yield valuable information about the magnetic structure of materials
provided one has a reliable theory for their interpretation.
Therefore,
the theoretical analysis of the dynamic quantities for quantum spin chains
is significant both from theoretical/academic and experimental/practical points of view.

Since the early 1930s the Bethe ansatz has been known as a powerful method of exploring quantum spin chains.
However,
only recently it has become possible to calculate quantities such as norms of
and matrix elements between Bethe ansatz states which are necessary to calculate dynamic quantities.
For recent Bethe ansatz results on the ground-state dynamic structure factors of the spin-1/2 $XXZ$ Heisenberg chain
see Refs. \onlinecite{01,02,03}.
We also mention here the field-theoretical approaches
for evaluation of the dynamic quantities \cite{04}
which, however, are restricted to low-energy physics only
and therefore can only give the dynamic quantities
in a small part of the plane wave-vector $\kappa$ - frequency $\omega$
(hereinafter the $\kappa$-$\omega$ plane).
Traditionally,
those calculations were performed for the Tomonaga-Luttinger model,
which describes one-dimensional spinless fermions moving in a continuum,
with linear dispersion relation.
Recently, however, the curvature of the dispersion relation has been taken into account
in calculating the properties of quantum wires \cite{curvature}.
The spin-1/2 $XXZ$ chain is a lattice system closely related to these continuum models.
The low-energy and long-wavelength limit of its ground-state $zz$ dynamic structure factor
was recently studied by combining several analytic and numeric techniques \cite{pereira}.
Recently \cite{skos} an extension of the density-matrix renormalization group (DMRG) method was proposed
which allows for the calculation of real-time correlation functions of $XXZ$ chains
at arbitrary finite temperatures in the thermodynamic limit.
However, numerical limitations presently restrict the time range over which results are reliable
to values comparable to those reached in complete diagonalization studies \cite{fls97}.

Another exactly solvable class of quantum spin chains are spin-1/2 $XY$ chains.
Rigorous analysis of these systems is based on exploiting the Jordan-Wigner transformation
to spinless fermions \cite{05}.
(For a relation between the Bethe ansatz method and the Jordan-Wigner approach
for the spin-1/2 $XX0$ (i.e. isotropic $XY$) chain see Ref. \onlinecite{06}.)
Although after applying the Jordan-Wigner transformation to the spin-1/2 $XY$ chains
one faces a system of noninteracting spinless fermions
the calculation of the spin correlation functions is not a trivial problem
because of the nonlocal character of the transformation.
Thus,
the $zz$ spin correlations are related to the two-fermion (density-density) correlations,
whereas,
e.g.,
the $xx$ spin correlations are related to many-fermion correlations.
Accordingly,
the $zz$ dynamics is well studied \cite{07,08},
whereas closed-form expressions,
e.g., for the $xx$ dynamic quantities are rather scarce \cite{09,10,11}
(see also references in Ref. \onlinecite{12}).

In the present paper we consider several quantum spin chains
with regular alternation or random disorder in the nearest-neighbor interactions
and follow the effect of such modifications
on the dynamic structure factors.
The inhomogeneity introduced refers mainly to the sign of interactions
and may mimic the ferromagnetic or antiferromagnetic types of nearest-neighbor exchange coupling.
The interest in models of such a kind is not purely theoretical.
Recently some organic and inorganic magnets have been recognized as
alternating sign \cite{stone}, random bond \cite{zheludev}
and alternating random bond \cite{manaka,hida,nakamura} antiferromagnetic spin chains.
The dynamic study of the quantum spin chain material with bond randomness
BaCu$_2$(Si$_{1-x}$Ge$_x$)$_2$O$_7$, $x=0.5$
using inelastic neutron scattering
revealed that its dynamic structure factor
can be fitted by the M\"{u}ller ansatz \cite{08} surprisingly well \cite{zheludev}.
The correspondence between the dynamic properties
of the random-bond Heisenberg antiferromagnetic spin chain
and the BaCu$_2$(Si$_{1-x}$Ge$_x$)$_2$O$_7$ compound
has been confirmed numerically by the quantum Monte-Carlo method \cite{24}.

In our calculation of dynamic quantities
we use appropriate transformations
to eliminate the inhomogeneity from the spin Hamiltonian
arriving at the homogenous model
the dynamic properties of which are well known.
Thus we reduce the complex behavior of dynamic quantities
for periodic/random quantum spin chains
to the known dynamic properties of the homogenous model.
In what follows
we deal with spin-1/2 isotropic $XY$ ($XX$ or $XX0$) chains
since the dynamic quantities for the more general case of the $XXZ$ Heisenberg exchange interaction
are less known.

The paper is organized as follows.
To the end of this section we introduce the spin model, the quantities of interest
and recall some results for the dynamic quantities
obtained within the Jordan-Wigner fermionization approach
which are used in the following sections.
In Sec.~\ref{sec2} we consider the spin-1/2 $XX$ chain
with regularly alternating or random sign of the $XX$ exchange interaction.
In Sec.~\ref{sec3} we consider the spin-1/2 $XX$ chain
with the Dzyaloshinskii-Moriya interaction
the sign of which may either vary regularly along the chain
(or it has a regularly varying component in addition to a constant component)
or may acquire its sign randomly.
We summarize our findings in Sec.~\ref{sec4}.

We consider the following Hamiltonian of a one-dimensional spin $s=1/2$  $XX$ model
with two-site interactions
which can be examined rigorously within the framework of the Jordan-Wigner approach \cite{05}:
\begin{eqnarray}
\label{1.01}
H=\sum_{n}
\left(J_n\left(s_n^x s_{n+1}^x + s_n^y s_{n+1}^y\right)
+ D_n\left(s_n^x s_{n+1}^y - s_n^y s_{n+1}^x\right)
+ \Omega s_n^z\right).
\end{eqnarray}
Here $J_n$ is the exchange $XX$ interaction between neighboring sites $n$ and $n+1$,
$D_n$ is the $z$-component of the Dzyaloshinskii-Moriya interaction between these sites,
and $\Omega$ is the external transverse ($z$) magnetic field.
The sum in (\ref{1.01}) runs over all $N$ sites;
the boundary conditions (periodic or open) are not essential
for the quantities considered below
which we calculate in the thermodynamic limit $N\to\infty$.

We are interested in the dynamic structure factors of the spin model (\ref{1.01})
(defined most conveniently for periodic boundary conditions,
so that $m=N$ is equivalent to $m=0$)
\begin{eqnarray}
\label{1.02}
S_{\alpha\beta}(\kappa,\omega)
=
\frac{1}{N}\sum_{j=1}^{N}
\sum_{m=1}^{N}\exp\left(-{\rm{i}}\kappa m\right)
\int_{-\infty}^{\infty}{\rm{d}}t\exp\left({\rm{i}}\omega t\right)
\left(\langle s_j^\alpha(t)s_{j+m}^\beta\rangle
-\langle s_j^\alpha\rangle\langle s_{j+m}^\beta\rangle\right),
\end{eqnarray}
where $\alpha,\beta=x,y,z$.
These experimentally accessible quantities contain important information about the spin model (\ref{1.01}).
By symmetry arguments
$S_{xx}(\kappa,\omega)=S_{yy}(\kappa,\omega)$,
$S_{xy}(\kappa,\omega)=-S_{yx}({-\kappa},\omega)$.
Therefore,
in what follows we may focus only on
$S_{xx}(\kappa,\omega)$, $S_{xy}(\kappa,\omega)$ and $S_{zz}(\kappa,\omega)$.
Moreover,
the model (\ref{1.01}) implies that
$\langle s_n^x\rangle=\langle s_n^y\rangle=0$
and hence the second term in the parentheses in Eq. (\ref{1.02}) may be omitted
if $\alpha,\beta=x,y$.

Consider first a uniform chain (\ref{1.01}) with $J_n=J$, $D_n=0$.
Again by symmetry arguments
$S_{zz}(\kappa,\omega)$ is insensitive to a sign change of the exchange interaction
$J\to -J$
whereas
$S_{xx}(\kappa,\omega)\to S_{xx}(\kappa\mp\pi,\omega)$,
$S_{xy}(\kappa,\omega)\to S_{xy}(\kappa\mp\pi,\omega)$.
Next,
from Refs. \onlinecite{07,08} we know that
\begin{eqnarray}
\label{1.03}
S_{zz}(\kappa,\omega)
=
\int_{-\pi}^{\pi}{\rm{d}}\kappa_1 n_{\kappa_1}\left(1-n_{\kappa+\kappa_1}\right)
\delta\left(\omega+\Lambda_{\kappa_1}-\Lambda_{\kappa+\kappa_1}\right),
\end{eqnarray}
where $\Lambda_\kappa=\Omega+J\cos\kappa$ is the elementary excitation energy
of the Jordan-Wigner fermions
and
$n_\kappa=1/\left(1+\exp\left(\beta\Lambda_\kappa\right)\right)$ is the Fermi function.
Obviously the $zz$ dynamic structure factor (\ref{1.03})
is governed by a continuum of two-fermion (particle-hole) excitations \cite{08}.
Let us introduce the following characteristic lines in the $\kappa$-$\omega$ plane
\begin{eqnarray}
\label{1.04}
\frac{\omega^{(1)}(\kappa)}{\vert J\vert}
=
2\left\vert\sin\frac{\kappa}{2}\sin\left(\frac{\vert\kappa\vert}{2}-\alpha\right)\right\vert,
\nonumber\\
\frac{\omega^{(2)}(\kappa)}{\vert J\vert}
=
2\left\vert\sin\frac{\kappa}{2}\sin\left(\frac{\vert\kappa\vert}{2}+\alpha\right)\right\vert,
\nonumber\\
\frac{\omega^{(3)}(\kappa)}{\vert J\vert}
=
2\left\vert\sin\frac{\kappa}{2}\right\vert,
\end{eqnarray}
where $\alpha=\arccos\left(\vert\Omega\vert/\vert J\vert\right)$
varies from $\pi/2$ (when $\Omega=0$)
to 0 (when $\vert\Omega\vert=\vert J\vert$).
The ground-state $S_{zz}(\kappa,\omega)$ is nonzero for $\vert \Omega\vert <\vert J\vert$
and in a restricted region in the $\kappa$-$\omega$ plane
(we assume $\vert\kappa\vert\le\pi$, $\omega\ge 0$)
with the lower boundary $\omega_l(\kappa)=\omega^{(1)}(\kappa)$
and the upper boundary
$\omega_u(\kappa)=\omega^{(2)}(\kappa)$ if $0\le\vert\kappa\vert\le\pi-2\alpha$
or
$\omega_u(\kappa)=\omega^{(3)}(\kappa)$ if $\pi-2\alpha\le\vert\kappa\vert\le\pi$.
Moreover,
$S_{zz}(\kappa,\omega)$ exhibits a finite jump
(increasing its value by 2) along the middle boundary
$\omega_m(\kappa)=\omega^{(2)}(\kappa)$,
$\pi-2\alpha\le\vert\kappa\vert\le\pi$.
Finally,
$S_{zz}(\kappa,\omega)$ shows a van Hove singularity along the curve
$\omega_s(\kappa)=\omega^{(3)}(\kappa)$.
As temperature increases
the lower boundary becomes smeared out and finally disappears.
The upper boundary is given by $\omega^{(3)}(\kappa)$
and $S_{zz}(\kappa,\omega)$ becomes field-independent in the high-temperature limit.

The $xx$/$xy$ dynamic structure factor is governed by many-fermion excitations
and therefore is a much more complicated quantity
(the two-fermion contribution to $S_{xx}(\kappa,\omega)$ was discussed in Refs. \onlinecite{13,14}).
However, the ground-state $S_{xx}(\kappa,\omega)$ and $S_{xy}(\kappa,\omega)$
can be easily calculated for strong fields $\vert\Omega\vert>\vert J\vert$ \cite{11}
\begin{eqnarray}
\label{1.05}
S_{xx}(\kappa,\omega)
={\rm{i}} \, {\rm{sgn}}(\Omega)S_{xy}(\kappa,\omega)
=\frac{\pi}{2}\delta\left(\omega-\vert\Omega\vert-J\cos\kappa\right).
\end{eqnarray}
Eq. (\ref{1.05}) shows that
all the spectral weight in this case is concentrated along the curve
\begin{eqnarray}
\label{1.06}
\frac{\omega^{\star}(\kappa)}{\vert J\vert}
=\frac{\vert\Omega\vert}{\vert J\vert}
+{\rm{sgn}}(J)\cos\kappa.
\end{eqnarray}
At sufficiently low temperatures ($k_{{\rm{B}}}T/\vert J\vert=0.01\ldots 0.05$) we know from numerics
(see Ref. \onlinecite{12})
that although
$S_{xx}(\kappa,\omega)$ and $S_{xy}(\kappa,\omega)$
are not {\it a priori} restricted to a certain region in the $\kappa$-$\omega$ plane
(and indeed these quantities have nonzero values throughout the $\kappa$-$\omega$ plane),
nevertheless their values are rather small outside the two-fermion excitation continuum discussed above.
More precisely,
the $xx$ and $xy$ dynamic structure factors show washed-out excitation branches
roughly following the boundaries of the two-fermion excitation continuum
(see Eq. (\ref{1.04})) for $J<0$
or following these boundaries shifted along the $\kappa$-axis by $\pi$ for $J>0$.
In the high-temperature limit we have \cite{09,10}
\begin{eqnarray}
\label{1.07}
S_{xx}(\kappa,\omega)
=\frac{\sqrt{\pi}}{4\vert J\vert}
\left(
\exp\left(-\frac{\left(\omega-\Omega\right)^2}{J^2}\right)
+
\exp\left(-\frac{\left(\omega+\Omega\right)^2}{J^2}\right)
\right),
\nonumber\\
{\rm{i}}S_{xy}(\kappa,\omega)
=\frac{\sqrt{\pi}}{4\vert J\vert}
\left(
\exp\left(-\frac{\left(\omega-\Omega\right)^2}{J^2}\right)
-
\exp\left(-\frac{\left(\omega+\Omega\right)^2}{J^2}\right)
\right),
\end{eqnarray}
i.e. the $xx$ and $xy$ dynamic structure factors in this case are $\kappa$-independent
and display Gaussian ridges at $\omega=\pm\Omega$.

Similar results on the dynamic properties of the dimerized spin-1/2 $XX$ chain
(i.e. with $J_n=J\left(1-(-1)^n\delta\right)$,
where $0<\delta<1$ is the dimerization parameter,
and $D_n=0$ in Eq. (\ref{1.01}))
can be found in Ref. \onlinecite{15} (and references therein).
The dynamic properties of the uniform spin-1/2 $XX$ chain with the Dzyaloshinskii-Moriya interaction
(i.e. with $J_n=J$,
$D_n=D$ in Eq. (\ref{1.01}))
were discussed in Ref. \onlinecite{16}.

In what follows (Sec.~\ref{sec2} and Sec.~\ref{sec3}) we use the results recalled here
to examine the dynamic properties of quantum spin chains
with special types of periodically varying or randomly distributed interspin interactions.

\section{Spin-1/2 $XX$ chain with periodicity/randomness
in the sign of exchange interaction}
\label{sec2}
\setcounter{equation}{0}

In this section we consider the spin model with the Hamiltonian (\ref{1.01})
assuming $J_n=\lambda_n J$ with $\lambda_n=\pm 1$ and $D_n=0$,
i.e. the exchange interaction between the sites $n$ and $n+1$
may be either antiferromagnetic if $\lambda_n J>0$
or ferromagnetic if $\lambda_n J<0$
depending on the given sequence $\{\lambda_1,\ldots,\lambda_N\}$.
Let us perform a gauge transformation
\begin{eqnarray}
\label{2.01}
s_n^x\to \tilde{s}_n^x=\lambda_1\lambda_2\ldots\lambda_{n-1}s_n^x,
\nonumber\\
s_n^y\to \tilde{s}_n^y=\lambda_1\lambda_2\ldots\lambda_{n-1}s_n^y,
\nonumber\\
s_n^z\to \tilde{s}_n^z=s_n^z
\end{eqnarray}
after which the Hamiltonian $H$ transforms into the Hamiltonian $\tilde{H}$
of the homogeneous model
with exchange constant $J_n\equiv J$
(up to an inessential boundary term).
(We denote the quantities related to the transformed (homogeneous) model by a tilde.)
Obviously,
according to (\ref{2.01})
the $zz$ dynamic structure (as well as all thermodynamic quantities)
does not feel an inhomogeneous sequence of signs
$\{\lambda_1,\ldots,\lambda_N\}$.
In contrast,
the $xx$ and $xy$ dynamic structure factors do depend on $\{\lambda_1,\ldots,\lambda_N\}$.
Below we consider separately the cases of periodic sequences
and of random sequences of signs.

\subsection{Periodic case}

We begin with the case of period $p=2$,
i.e. $\{\lambda_n\}=\{1,-1,1,-1,\ldots\}$.
After performing the transformation (\ref{2.01})
we have
$\tilde{s}^{\alpha}_{2j-1}=(-1)^{j+1}s^{\alpha}_{2j-1}$,
$\tilde{s}^{\alpha}_{2j}=(-1)^{j+1}s^{\alpha}_{2j}$,
$j=1,2,\ldots$
(here and to the end of the paper
$\alpha,\beta=x,y$)
and therefore according to (\ref{1.02}) we can write
\begin{eqnarray}
\label{2.02}
S_{\alpha\beta}(\kappa,\omega)
=
\frac{1}{2}\sum_{m=1}^{N}\exp\left(-{\rm{i}}\kappa m\right)
\int_{-\infty}^{\infty}{\rm{d}}t\exp\left({\rm{i}}\omega t\right)
a_m
\langle \tilde{s}_1^{\alpha}(t)\tilde{s}_{1+m}^{\beta}\rangle
\nonumber\\
+\frac{1}{2}\sum_{m=1}^{N}\exp\left(-{\rm{i}}\kappa m\right)
\int_{-\infty}^{\infty}{\rm{d}}t\exp\left({\rm{i}}\omega t\right)
b_m
\langle \tilde{s}_2^{\alpha}(t)\tilde{s}_{2+m}^{\beta}\rangle,
\end{eqnarray}
where
$\{a_1,a_2,a_3,\ldots\}=\{1,-1,-1,1,1,-1,-1,\ldots\}$,
$\{b_1,b_2,b_3,\ldots\}=\{-1,-1,1,1,-1,-1,1,\ldots\}$.
Noting that
$a_m=((1-{\rm{i}})/2)\exp({\rm{i}}\pi m/2)
+((1+{\rm{i}})/2)\exp(3{\rm{i}}\pi m/2)$
and $b_m=a_{m+1}$ we immediately find from Eq. (\ref{2.02}) that
\begin{eqnarray}
\label{2.03}
S_{\alpha\beta}(\kappa,\omega)
=
\frac{1}{2}\tilde{S}_{\alpha\beta}\left(\kappa+\frac{\pi}{2},\omega\right)
+
\frac{1}{2}\tilde{S}_{\alpha\beta}\left(\kappa+\frac{3\pi}{2},\omega\right).
\end{eqnarray}
On the l.h.s in Eq. (\ref{2.03}) we have the dynamic structure factors
for the periodic chain $S_{\alpha\beta}(\kappa,\omega)$
whereas on the r.h.s. in Eq. (\ref{2.03})
the dynamic structure factors $\tilde{S}_{\alpha\beta}(\kappa,\omega)$
refer to the uniform chain with the exchange constant $J$;
the latter quantities were discussed in Sec.~\ref{sec1}.
These calculations can be easily extended for periodic chains of larger periods.
For example,
for $p=3$ with $\{\lambda_n\}=\{1,1,-1,1,1,-1,\ldots\}$
after performing similar calculations
we arrive instead of Eq. (\ref{2.03}) at
\begin{eqnarray}
\label{2.04}
S_{\alpha\beta}(\kappa,\omega)
=
\frac{4}{9}\tilde{S}_{\alpha\beta}\left(\kappa+\frac{\pi}{3},\omega\right)
+
\frac{1}{9}\tilde{S}_{\alpha\beta}\left(\kappa+\pi,\omega\right)
+
\frac{4}{9}\tilde{S}_{\alpha\beta}\left(\kappa+\frac{5\pi}{3},\omega\right).
\end{eqnarray}

\begin{figure}
\epsfig{file = 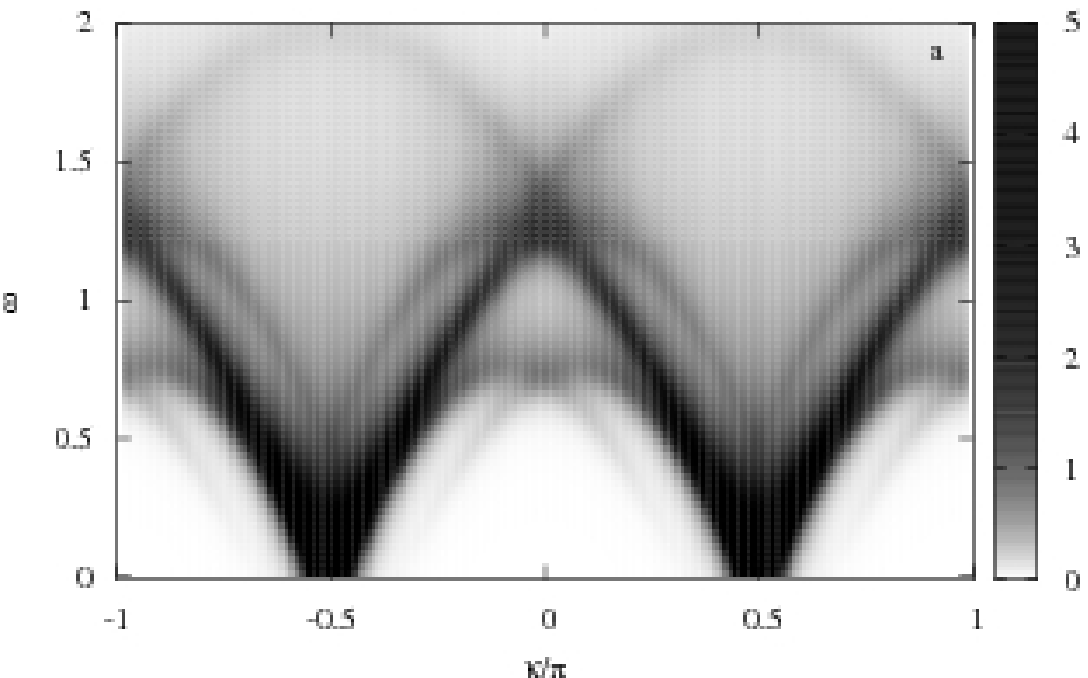, height = 0.25\linewidth}\\
\epsfig{file = 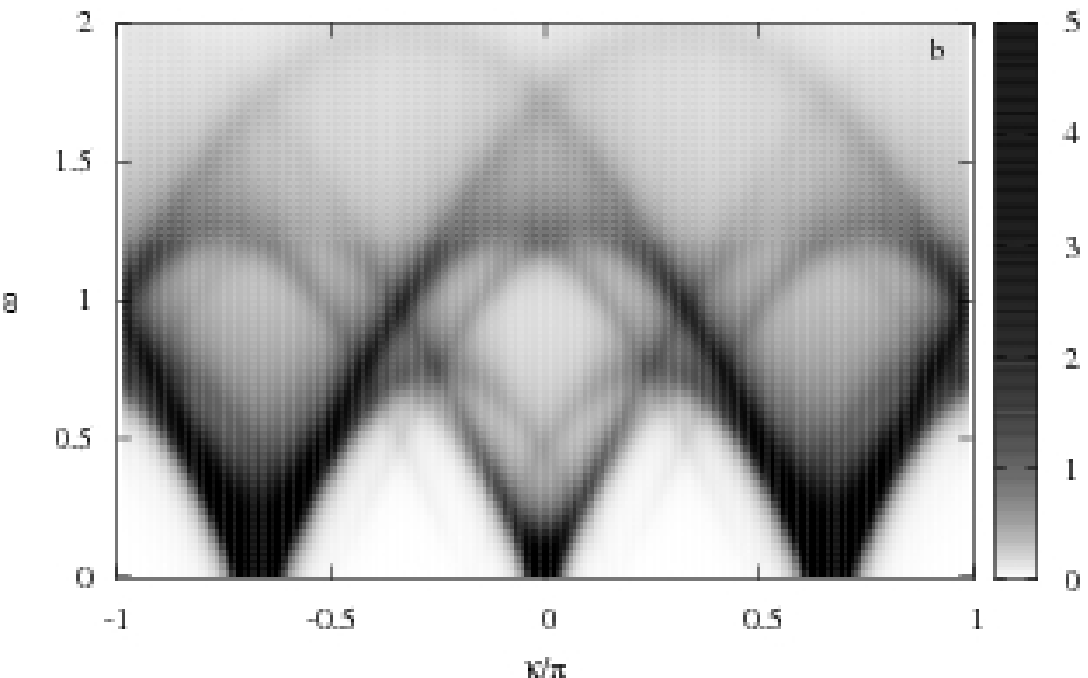, height = 0.25\linewidth}\\
\epsfig{file = 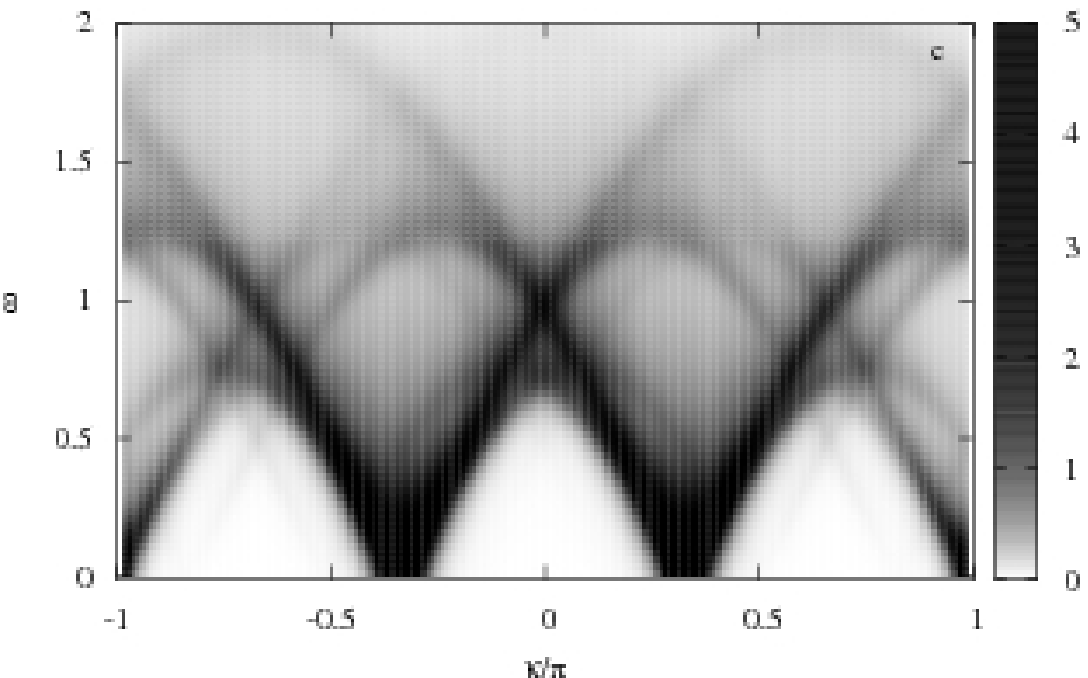,  height = 0.25\linewidth}\\
\caption{$S_{xx}(\kappa,\omega)$
for the spin-1/2 $XX$ chain
with periodic sequences of exchange interactions
$\{J,-J,J,-J,\ldots\}$, $J=1$ (upper panel a)
and
$\{J,J,-J,J,J,-J,\ldots\}$
with $J=1$ (middle panel b) and with $J=-1$ (lower panel c)
for $\Omega=0.25$
at low temperature, $\beta =20$.}
\label{fig1}
\end{figure}

To illustrate the effect of a regularly alternating sign of exchange interaction on $S_{xx}(\kappa,\omega)$
we display this quantity
calculated according to Eqs. (\ref{2.03}), (\ref{2.04})
in Fig.~\ref{fig1}.
Evidently,
in the high-temperature limit owing to $\kappa$-independence of $S_{xx}(\kappa,\omega)$
(see Eq. (\ref{1.07}))
regular alternation of the exchange interaction signs does not manifest itself
in the $xx$ dynamic structure factor.
However at low temperatures it may lead to rather intricate frequency/wave-vector patterns
(see Fig.~\ref{fig1}).
Interestingly,
we may reproduce the sequence $\{\lambda_n\}$
knowing the number of soft modes $\kappa_0$ and their position.
In the limit $T=0$ and $\vert\Omega\vert>\vert J\vert$
we may insert Eq. (\ref{1.05}) into the r.h.s. of Eqs. (\ref{2.03}), (\ref{2.04})
to find that the spectral weight is concentrated along the curves
which follow from Eq. (\ref{1.06})
after corresponding shifts along the $\kappa$-axis.

The results of this subsection
are complementary to the earlier results on thermodynamic and dynamic properties
of periodic spin-1/2 $XX$ chains (see Refs. \onlinecite{17,18,19} and references therein).

\subsection{Random case}

We now proceed with the case of randomly distributed signs of exchange interactions
assuming $\{\lambda_n\}$ to be a sequence of independent random variables
each with the following bimodal probability distribution
\begin{eqnarray}
\label{2.05}
p(\lambda_n)=p\delta(\lambda_n+1)+(1-p)\delta(\lambda_n-1),
\end{eqnarray}
where $0\leq p\leq 1$.
We are interested in random-averaged quantities
and denote the average over all realizations of randomness as
$\overline{(\ldots)}
=\prod_n\int_{-\infty}^{\infty}{\rm{d}}\lambda_n p(\lambda_n)(\ldots)$.
Random chains of that type (in fact, for more general $XXZ$ coupling) were studied
in Refs. \onlinecite{20,21}.

\begin{figure}
\epsfig{file = 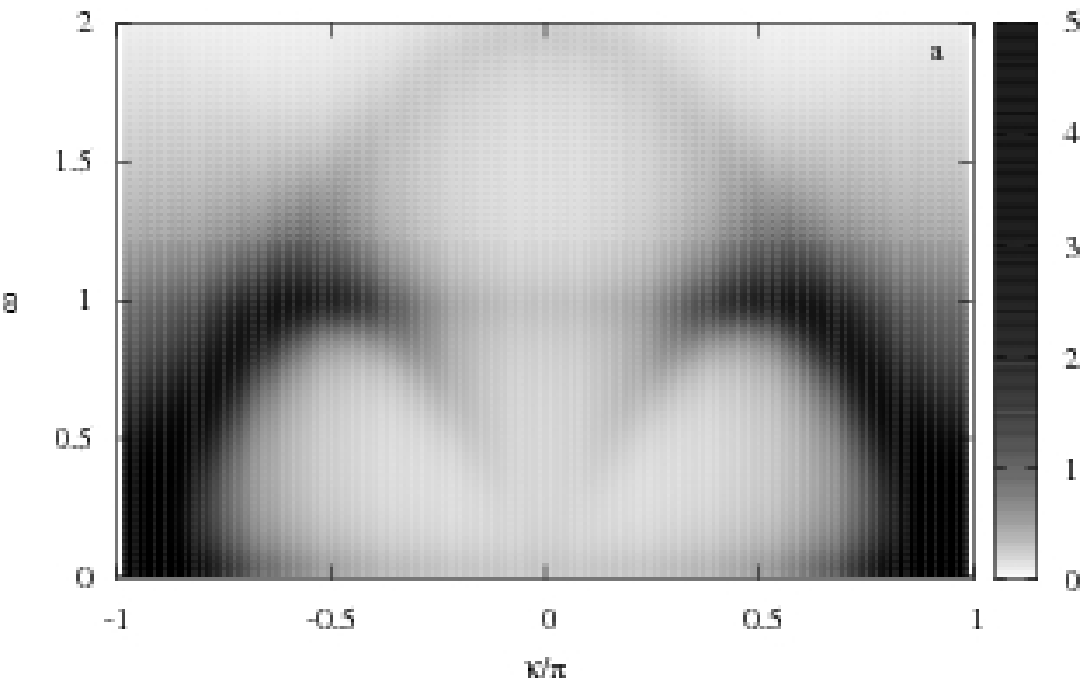, height = 0.25\linewidth}
\epsfig{file = 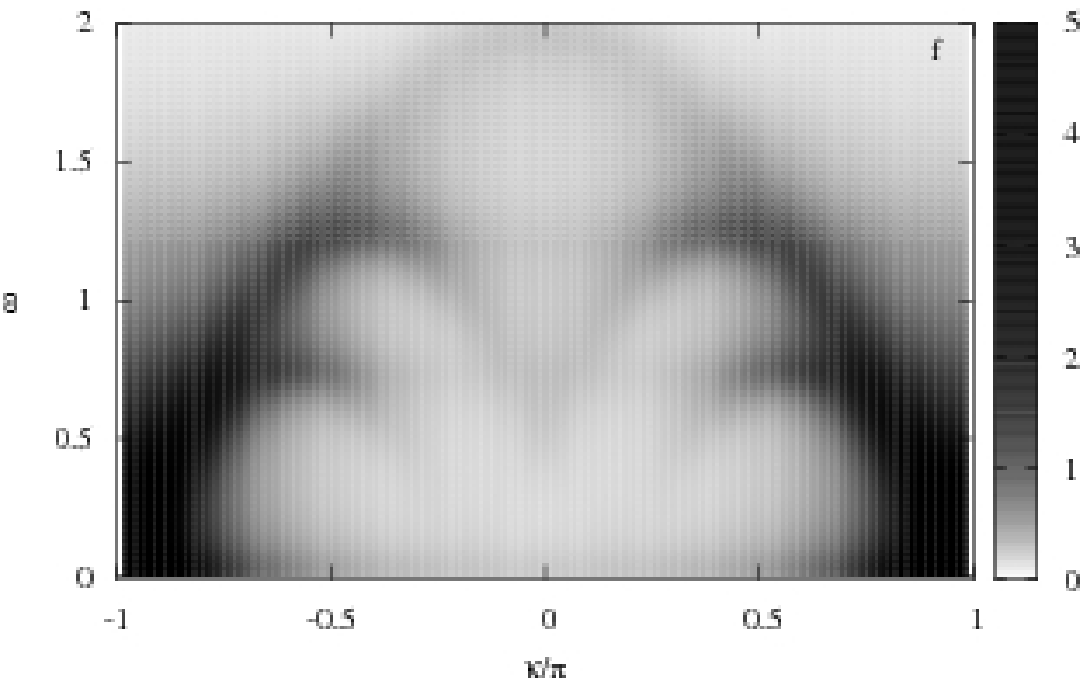, height = 0.25\linewidth}\\
\epsfig{file = 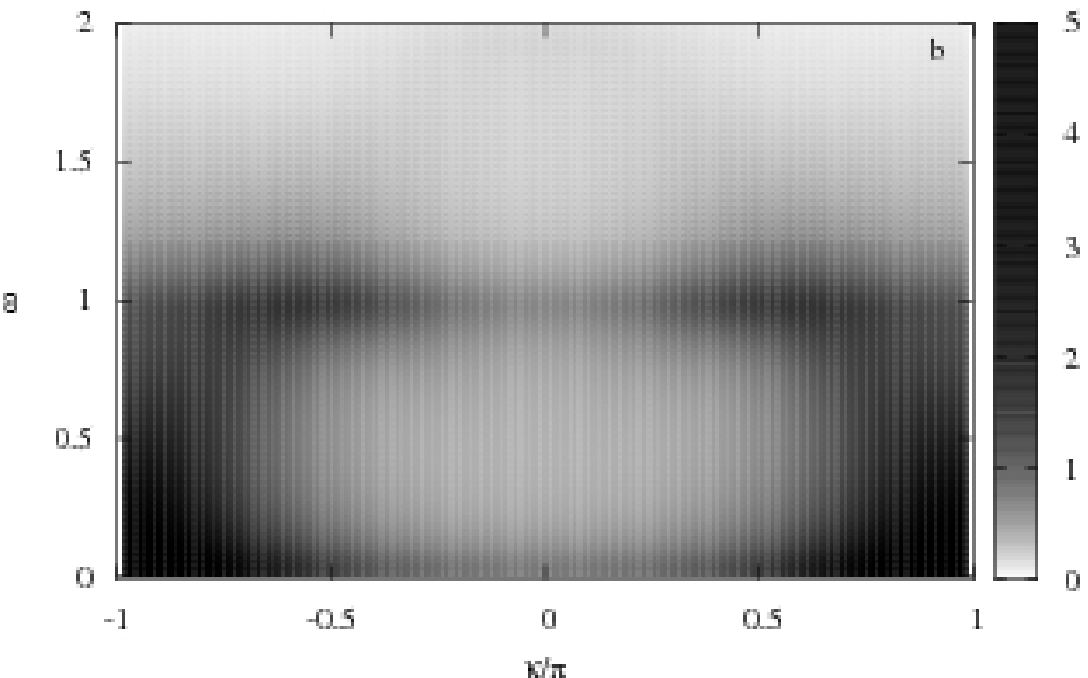, height = 0.25\linewidth}
\epsfig{file = 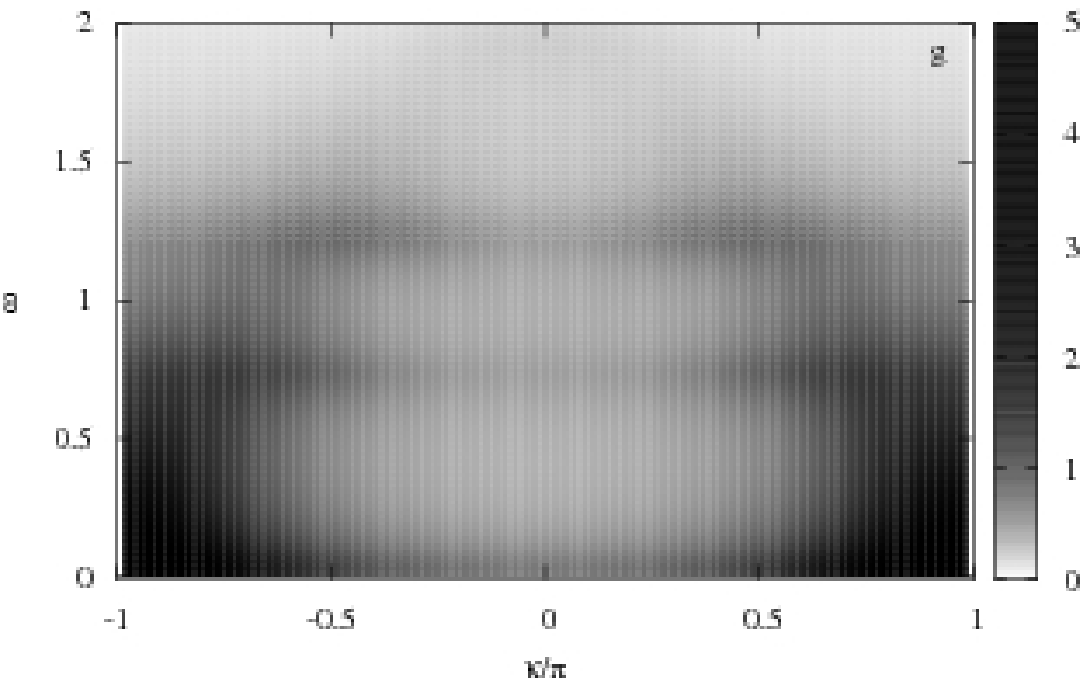, height = 0.25\linewidth}\\
\epsfig{file = 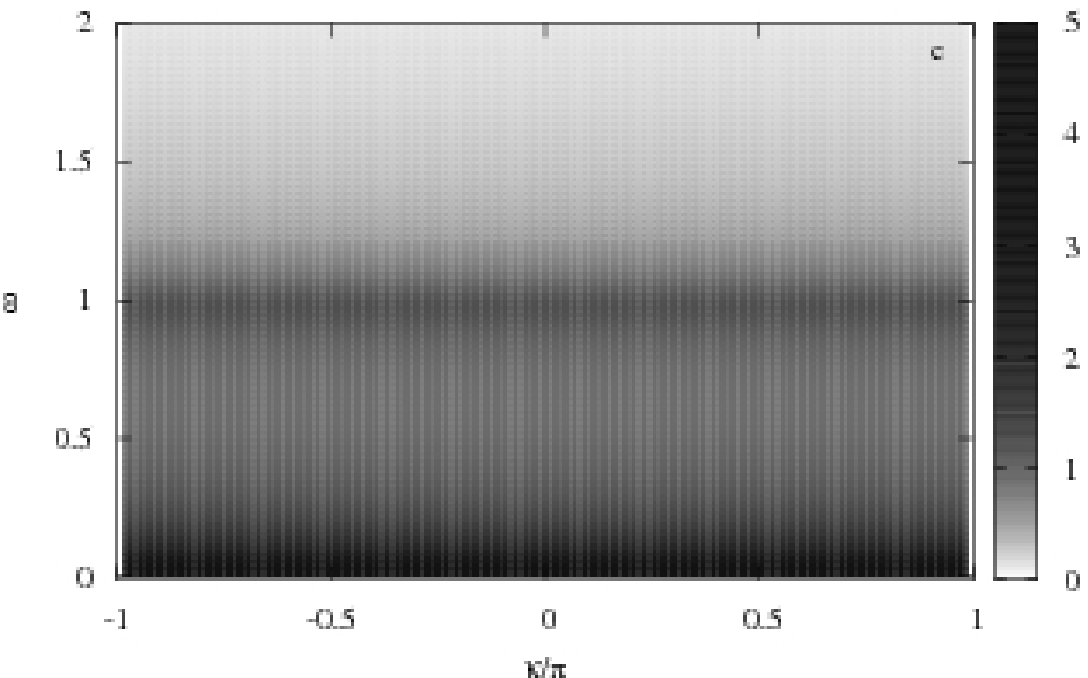, height = 0.25\linewidth}
\epsfig{file = 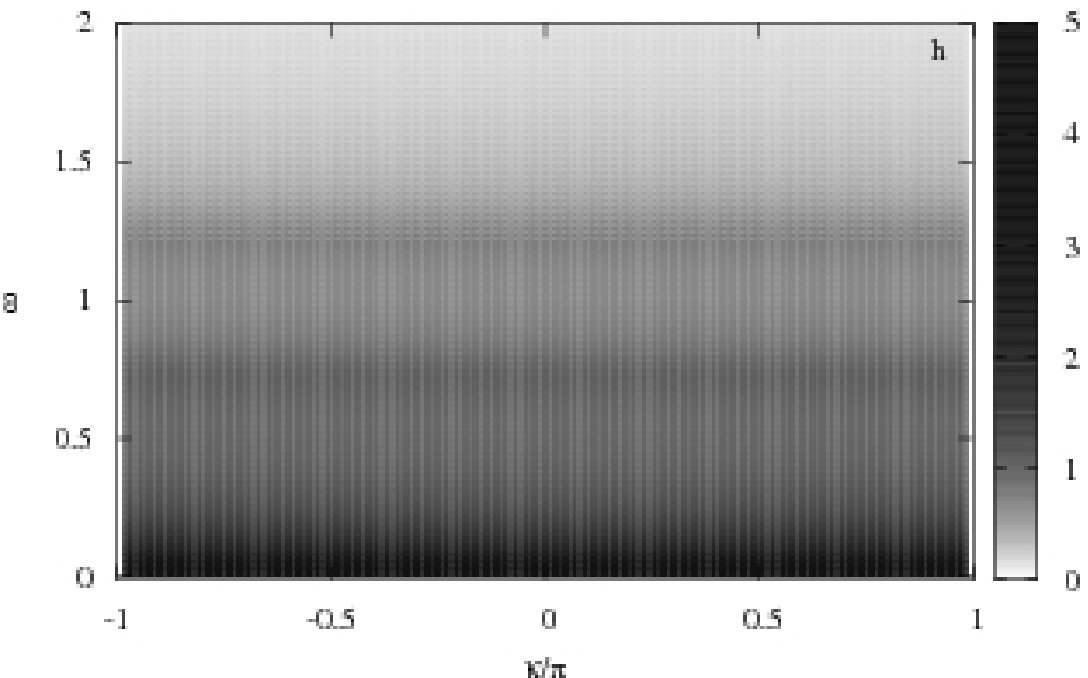, height = 0.25\linewidth}\\
\epsfig{file = 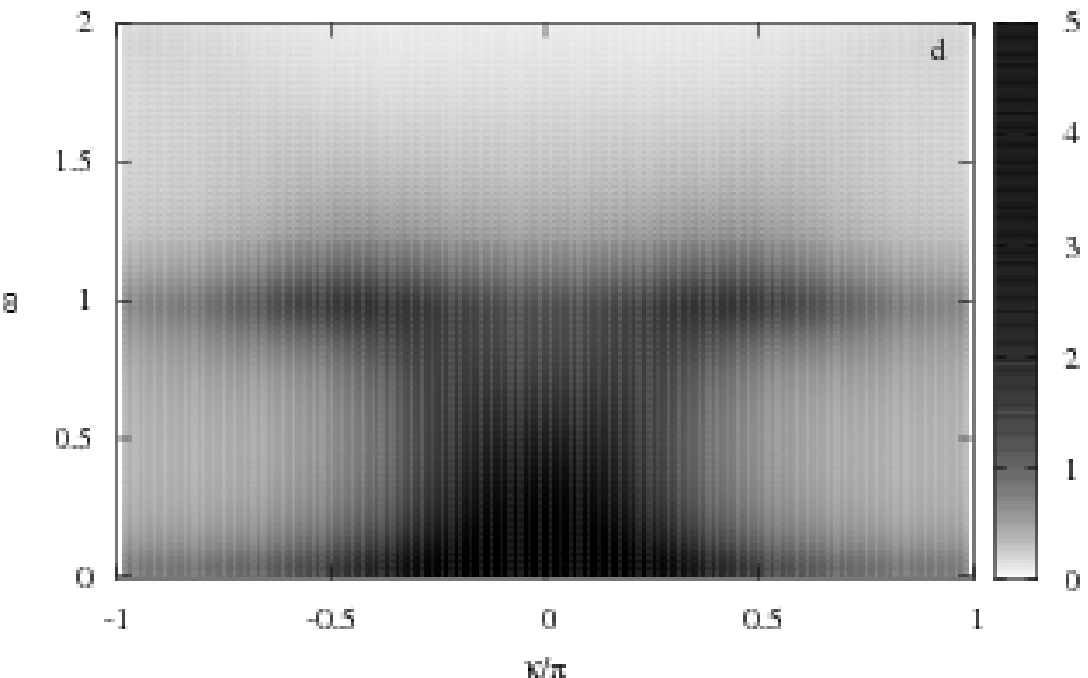, height = 0.25\linewidth}
\epsfig{file = 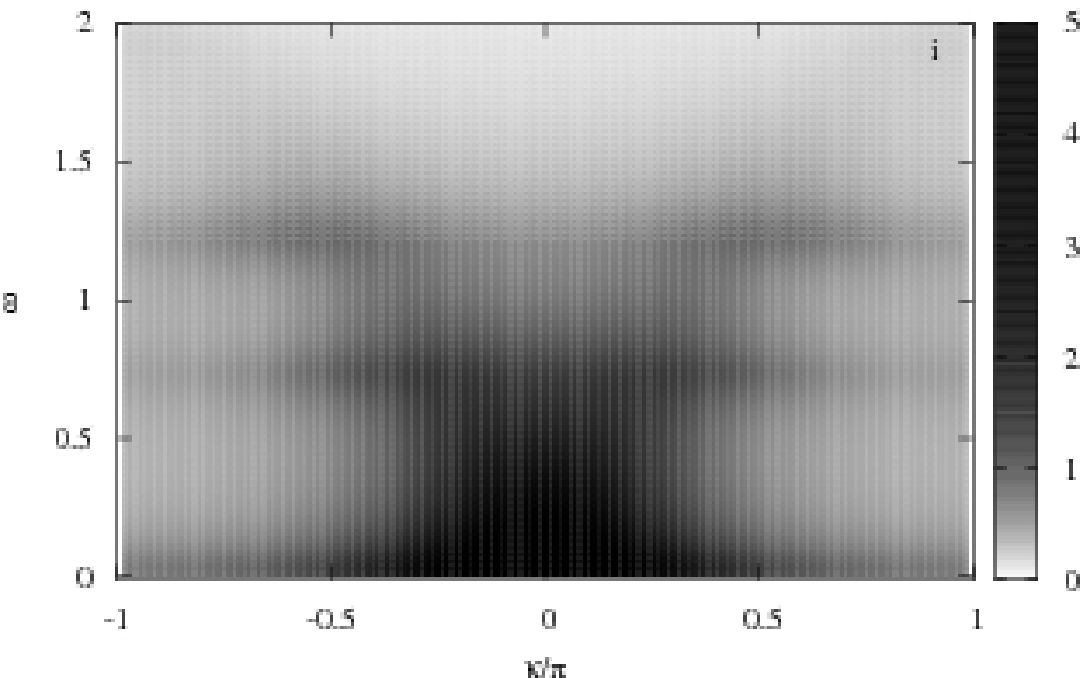, height = 0.25\linewidth}\\
\epsfig{file = 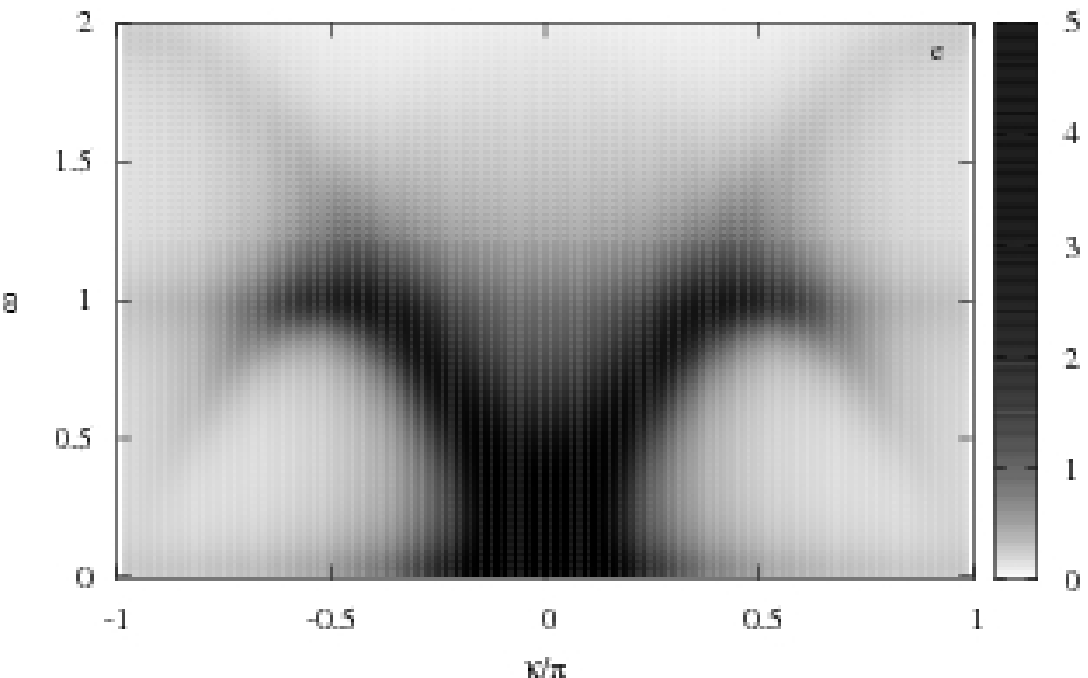, height = 0.25\linewidth}
\epsfig{file = 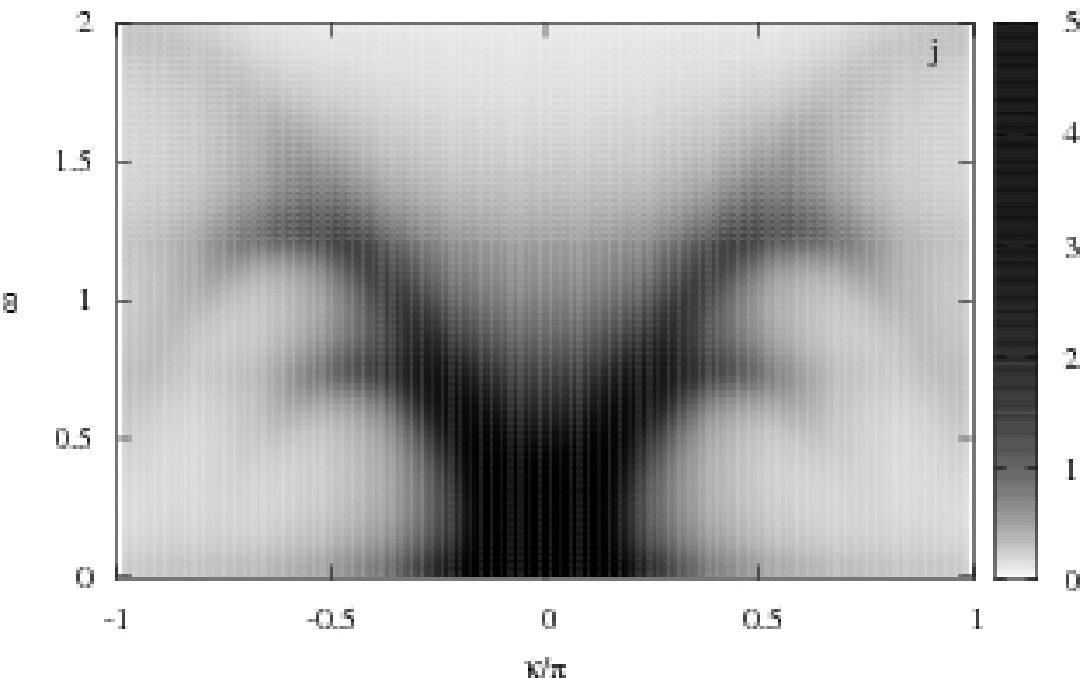, height = 0.25\linewidth}
\caption{$\overline{S_{xx}(\kappa,\omega)}$
for the spin-1/2 $XX$ chain with random-sign exchange interaction $J=1$,
$\Omega=0.01$ (left panels, a \ldots e),
$\Omega=0.25$ (right panels, f \ldots j),
$\beta=20$.
The values of $p$ in Eq. (\ref{2.05}) are as follows:
from top to bottom
$p=0.1$ ($\xi\approx 4.48$),
$p=0.25$ ($\xi\approx 1.44$),
$p=0.5$ ($\xi=0$),
$p=0.75$ ($\xi\approx 1.44$),
$p=0.9$ ($\xi\approx 4.48$).}
\label{fig2}
\end{figure}

Exploiting the gauge transformation (\ref{2.01}) and Eq. (\ref{2.05}) we find
\begin{eqnarray}
\label{2.06}
\overline{\langle s_j^{\alpha}(t)s_{j+m}^{\beta}\rangle}
=
(1-2p)^{\vert m\vert}
\langle \tilde{s}_j^{\alpha}(t) \tilde{s}_{j+m}^{\beta}\rangle.
\end{eqnarray}
Introducing the correlation length $\xi=-1/\ln\vert 1-2p\vert$,
the last expression (\ref{2.06}) can be rewritten as
\begin{eqnarray}
\label{2.07}
\overline{\langle s_j^{\alpha}(t)s_{j+m}^{\beta}\rangle}
=
\left\{
\begin{array}{ll}
\exp\left(-\frac{\vert m\vert}{\xi}\right) \langle \tilde{s}_j^{\alpha}(t) \tilde{s}_{j+m}^{\beta}\rangle, &
0\leq p \leq \frac{1}{2},\\
(-1)^{m} \exp\left(-\frac{\vert m\vert}{\xi}\right) \langle\tilde{s}_j^{\alpha}(t)\tilde{s}_{j+m}^{\beta}\rangle, &
\frac{1}{2}\leq p \leq 1.
\end{array}
\right.
\end{eqnarray}
As a result,
the random-averaged dynamic structure factors (\ref{1.02})
can be written as follows
\begin{eqnarray}
\label{2.08}
\overline{S_{\alpha\beta}(\kappa,\omega)}
=
\sum_{m=0,\pm1,\pm2,\ldots}
\exp\left(-{\rm{i}}\kappa m-\frac{\vert m\vert}{\xi}\right)
\int_{-\infty}^{\infty}{\rm{d}}t\exp\left({\rm{i}}\omega t\right)
\langle \tilde{s}_j^{\alpha}(t) \tilde{s}_{j+m}^{\beta}\rangle;
\end{eqnarray}
here $0\leq p\leq 1/2$.
If $1/2 \leq p\leq 1$,
a factor $(-1)^m$
(see Eq. (\ref{2.07}))
should be taken into account in Eq. (\ref{2.08})
and the resulting expression
$\overline{S_{\alpha\beta}(\kappa,\omega)}$ for $1/2\leq p \leq 1$
corresponds to $\overline{S_{\alpha\beta}(\kappa\mp\pi,\omega)}$
in formula (\ref{2.08}).
We use Eq. (\ref{2.08}) to compute $\overline{S_{\alpha\beta}(\kappa,\omega)}$
through the known results
for $\langle \tilde{s}_j^{\alpha}(t) \tilde{s}_{j+m}^{\beta}\rangle$ for the uniform chain
with exchange constant $J$
obtained analytically or numerically \cite{22,12}
(see Fig.~\ref{fig2}).

Let us consider the case $T=0$, $\vert\Omega\vert >\vert J\vert$
when the $xx$ dynamic structure factor $\tilde{S}_{xx}(\kappa,\omega)$ is given by Eq. (\ref{1.05}).
In the site representation we have
\begin{eqnarray}
\label{2.09}
\tilde S_{xx}(m,\omega)
=\int_{-\infty}^{\infty}{\rm{d}}t\exp\left({\rm{i}}\omega t\right)
\langle \tilde{s}_{j}^{x}(t)\tilde s_{j+m}^{x}\rangle
=\frac{1}{N}\sum_{\kappa}\exp\left({\rm{i}}\kappa m\right)\tilde{S}_{xx}(\kappa,\omega)
\nonumber\\
=\frac{1}{4}\int_{-\pi}^{\pi}{\rm{d}}\kappa\exp\left({\rm{i}}\kappa m\right)
\delta\left(\omega-\vert\Omega\vert-J\cos\kappa\right)
\nonumber\\
=
\frac{\cos(m\kappa_0)}
{2\vert J\sin\kappa_0\vert}
\theta\left(\omega-\vert\Omega\vert+\vert J\vert\right)
\theta\left(\vert\Omega\vert+\vert J\vert-\omega\right),
\end{eqnarray}
where $\kappa_0=\arccos((\omega-\vert\Omega\vert)/J)$.
[In Eq. (\ref{2.09}) and Eqs. (\ref{2.10}), (\ref{2.11})
the $\theta$-functions simply indicate the frequency range within which
the equation $\omega-\vert\Omega\vert-J\cos\kappa=0$
has the solutions $\kappa=\pm\kappa_0$.]
Consider first the case $0\leq p\leq 1/2$.
After substitution of (\ref{2.09}) into (\ref{2.08}) and some simple calculations one finds
\begin{eqnarray}
\label{2.10}
\overline{S_{xx}(\kappa,\omega)}
=
\sum_{m=0,\pm1,\pm2,\ldots}
\exp\left(-{\rm{i}}\kappa m-\frac{\vert m\vert}{\xi}\right)
\tilde{S}_{xx}(m,\omega)
\nonumber\\
=
\frac{1}{2\vert J\sin\kappa_0\vert}
\sum_{m=0,\pm 1,\pm 2,\dots}
\exp\left(-{\rm{i}}\kappa m-\frac{\vert m\vert}{\xi}\right)
\cos(m\kappa_0)
\nonumber\\
\times
\theta\left(\omega-\vert\Omega\vert+\vert J\vert\right)
\theta\left(\vert\Omega\vert+\vert J\vert-\omega\right)
\nonumber\\
=
\frac{1}{2\sqrt{J^2-\left(\omega-\vert\Omega\vert\right)^2}}
\frac{J\sinh\frac{1}{\xi}\left(J\cosh\frac{1}{\xi}-\left(\omega-\vert\Omega\vert\right)\cos\kappa\right)}
{\left(\omega-\vert\Omega\vert-J\cosh\frac{1}{\xi}\cos\kappa\right)^2+J^2\sinh^2\frac{1}{\xi}\sin^2\kappa}
\nonumber\\
\times
\theta\left(\omega-\vert\Omega\vert+\vert J\vert\right)
\theta\left(\vert\Omega\vert+\vert J\vert-\omega\right).
\end{eqnarray}
If $1/2\leq p\leq 1$, $\overline{S_{xx}(\kappa,\omega)}$ follows from Eq. (\ref{2.10})
after the change $\kappa\to\kappa\mp \pi$.
One can easily note
that Eq. (\ref{2.10}) transforms into Eq. (\ref{1.05}) in the nonrandom limit $1/\xi\to 0$
(i.e. $p\to 0$ or $p\to 1$)
[to show this one has to exploit the relation
$\lim_{\Gamma\to +0}\left(\Gamma/\left(\left(\omega-\omega_0\right)^2+\Gamma^2\right)\right)
=\pi\delta\left(\omega-\omega_0\right)$].
In the opposite limit of a completely random system
$1/\xi\to\infty$ (i.e. $p\to 1/2$) Eq. (\ref{2.10}) becomes
\begin{eqnarray}
\label{2.11}
\overline{S_{xx}(\kappa,\omega)}
=
\frac{1}{2\sqrt{J^2-\left(\omega-\vert \Omega\vert\right)^2}}
\theta\left(\omega-\vert\Omega\vert+\vert J\vert\right)
\theta\left(\vert\Omega\vert+\vert J\vert-\omega\right).
\end{eqnarray}
One immediately recognizes
that Eq. (\ref{2.11}) contains the contribution of only the autocorrelation function
(as it should be since the correlation length $\xi$ tends to zero)
and since in the limit considered
$4\langle s_j^x(t)s_j^x\rangle
=(1/N)\sum_{\kappa}\exp\left(-{\rm{i}}\Lambda_{\kappa}t\right)$ \cite{11}
the $xx$ dynamic structure factor is proportional
to the density of states of elementary excitations
$\rho(E)=(1/N)\sum_{\kappa}\delta\left(E-\Lambda_{\kappa}\right)$,
i.e. $\overline{S_{xx}(\kappa,\omega)}=(\pi/2)\rho(\omega)$
independent of $\kappa$.
For other values of $p$
$\overline{S_{xx}(\kappa,\omega)}$ (\ref{2.10})
is restricted to the frequency region
$\vert\Omega\vert-\vert J\vert<\omega<\vert\Omega\vert+\vert J\vert$
and shows square-root singularities as
$\omega\to\vert\Omega\vert\pm\vert J\vert$.
The frequency profiles at fixed $\kappa$ resemble
(although are not identical to)
Lorentzian shapes
centered at $\omega=\vert\Omega\vert +J\cosh(1/\xi)\cos\kappa$
with the line width $\Gamma=\vert J\sinh(1/\xi)\sin\kappa\vert$.

For nonzero temperature, $T\ne 0$,
and for subcritical field values, $\vert\Omega\vert<J$,
(\ref{2.08}) must be evaluated numerically
(see Fig.~\ref{fig2}).
In the case $p=1/2$
the correlation length $\xi\to 0$ and one expects only the autocorrelation function to contribute
to the $\kappa$-independent $\overline{S_{xx}(\kappa,\omega)}$
and
the frequency shape for any $\kappa$ is determined by the $\omega$-dependence of
$\tilde{S}_{xx}(0,\omega)
=\int_{-\infty}^{\infty}{\rm{d}}t\exp\left({\rm{i}}\omega t\right)
\langle \tilde{s}_j^x(t) \tilde{s}_j^x\rangle$
($\kappa$-independent stripes near frequencies which dominate the autocorrelation function).

We note some similarities to recent numerical results
on the spin-1/2 Ising chain in a random transverse field \cite{23}.
In particular,
the horizontal ($\kappa$-independent) stripe-like patterns in Fig. \ref{fig2}
resemble the results of Ref. \onlinecite{23} for strong disorder.
This is to be expected since for strong enough disorder only local correlations survive
and lead to a $\kappa$-independent dynamic structure factor.

The scheme presented here
can be also easily adapted to more complex models
where alternation and randomness are mixed.
For example,
the ferromagnetic-antiferromagnetic random alternating quantum spin chain compound
(CH$_3$)$_2$CHNH$_3$Cu(Cl$_x$Br$_{1-x}$)$_3$ can be viewed
as a spin-1/2 random alternating quantum Heisenberg chain \cite{nakamura}
\begin{eqnarray}
\label{2.12}
H=\sum_n
\left(J_{2n-1}\vec{s}_{2n-1}\cdot\vec{s}_{2n}
+J_{2n}\vec{s}_{2n}\cdot\vec{s}_{2n+1}\right),
\end{eqnarray}
where $J_{2n-1}=J$ is the weak uniform exchange bond,
$J_{2n}=2\lambda_{2n}J$ is the strong random-sign exchange bond
and $\{\lambda_{2n}\}$ is the sequence of independent random variables
each with the bimodal probability distribution (\ref{2.05}).
If we restrict ourselves to isotropic $XY$ interactions between spins in (\ref{2.12}),
the randomness can be excluded from the Hamiltonian by a slightly modified
gauge transformation
$\tilde s_{2n-1}^\alpha=s_{2n-1}^\alpha\prod_{m=1}^{n-1}\lambda_{2m}$,
$\tilde s_{2n}^\alpha=s_{2n}^\alpha\prod_{m=1}^{n-1}\lambda_{2m}$,
$n=2,3\ldots$
obtaining finally the Hamiltonian of a dimerized $XX$ chain
with the periodically varying exchange couplings $J,2J,J,2J\ldots$.
The random-averaged dynamic structure factors can be calculated
analogously to (\ref{2.06}) -- (\ref{2.10}).

\section{Spin-1/2 $XX$ chain
with periodicity/randomness in the sign of Dzyaloshinskii-Moriya interaction}
\label{sec3}
\setcounter{equation}{0}

We now consider the spin model with the Hamiltonian (\ref{1.01})
assuming $J_n=J$ and $D_n=\lambda_n D$ with $\lambda_n=\pm 1$.
(We note that the case
$J_n=\lambda_n J$, $D_n=D$
may be analyzed on the basis of the results reported below
after exploiting the unitary transformation discussed in Ref. \onlinecite{25}.)
It is generally known \cite{26,27,28,29,16}
that the Dzyaloshinskii-Moriya interaction $D_n$
can be eliminated from the Hamiltonian $H$ (\ref{1.01})
(up to an inessential boundary term)
by the spin coordinate transformation
\begin{eqnarray}
\label{3.01}
s_n^x\to\tilde{s}_n^x=\cos\phi_n s_n^x + \sin\phi_n s_n^y,
\nonumber\\
s_n^y\to\tilde{s}_n^y=-\sin\phi_n s_n^x + \cos\phi_n s_n^y,
\nonumber\\
s_n^z\to\tilde{s}_n^z = s_n^z,
\end{eqnarray}
where $\phi_{n}=\sum_{m=0}^{n-1}\varphi_m$,
$\varphi_0$ is an arbitrary angle which is usually assumed to be zero
and
$\tan\varphi_m=D_m/J$, $m=1,2\ldots$.
As a result,
one faces the Hamiltonian $\tilde{H}$ (\ref{1.01}) without the Dzyaloshinskii-Moriya interaction,
however,
with a renormalized $XX$ exchange interaction
$\tilde{J}_n={\rm{sgn}}(J)\sqrt{J^2+D^2_n}$.
In the uniform case, when $D_n=D$,
the unitary transformation (\ref{3.01})
was used in the recent studies of dynamics of quantum spin chains \cite{28,29,16}.
In this section we consider separately the two cases of periodically varying Dzyaloshinskii-Moriya interaction
and of random-sign Dzyaloshinskii-Moriya interaction
focusing on the $xx$ dynamic structure factor $S_{xx}(\kappa,\omega)$.

\subsection{Periodic case}

We begin with the case $p=2$ with
$\{\lambda_n\}=\{1,-1,1,-1,\ldots\}$,
i.e. $D_n=(-1)^{n+1}D$.
Then we have to put in Eq. (\ref{3.01})
$\varphi_m=(-1)^{m+1}\varphi$,
$\varphi=\arctan(D/J)$.
Moreover, it is convenient to assume $\varphi_0=-\varphi/2$.
Then $\phi_n=(-1)^n\varphi/2$ and the inverse transformation to the one given by (\ref{3.01}) reads
\begin{eqnarray}
\label{3.02}
s_n^x=\cos\frac{\varphi}{2}\, \tilde{s}_n^x -(-1)^n \sin\frac{\varphi}{2}\,\tilde{s}_n^y,
\nonumber\\
s_n^y=(-1)^n\sin\frac{\varphi}{2}\, \tilde{s}_n^x + \cos\frac{\varphi}{2}\, \tilde{s}_n^y,
\nonumber\\
s_n^z=\tilde{s}_n^z.
\end{eqnarray}
By substituting Eq. (\ref{3.02}) into Eq. (\ref{1.02}) one immediately finds
that the $xx$ dynamic structure factor $S_{xx}(\kappa,\omega)$
of the $XX$ chain with the alternating Dzyaloshinskii-Moriya interaction $D,-D,D,-D,\ldots$
can be expressed through the $xx$ dynamic structure factor $\tilde{S}_{xx}(\kappa,\omega)$
of the uniform chain
with only $XX$ exchange interaction $\tilde{J}={\rm{sgn}}(J)\sqrt{J^2+D^2}$
as follows
\begin{eqnarray}
\label{3.03}
S_{xx}(\kappa,\omega)
=\cos^2\frac{\varphi}{2}\, \tilde{S}_{xx}(\kappa,\omega)
+\sin^2\frac{\varphi}{2}\, \tilde{S}_{xx}(\kappa\mp\pi,\omega).
\end{eqnarray}

We notice here that in the case when in Eq. (\ref{1.01}) $J_n=J$,  $D_n=D_0$,
that is, for uniform $XX$ and Dzyaloshinskii-Moriya couplings,
the relation for $S_{xx}(\kappa,\omega)$ is quite different:
$S_{xx}(\kappa,\omega)
=(\tilde{S}_{xx}(\kappa-\varphi,\omega)
+\tilde{S}_{xx}(\kappa+\varphi,\omega)
+{\rm{i}}\tilde{S}_{xy}(\kappa-\varphi,\omega)
-{\rm{i}}\tilde{S}_{xy}(\kappa+\varphi,\omega))/2$;
here $\tilde{S}_{\alpha\beta}(\kappa,\omega)$
is related to the uniform chain with only $XX$ exchange interaction
$\tilde{J}={\rm{sgn}}(J)\sqrt{J^2+D_0^2}$
(see Ref. \onlinecite{16}).
It is worth therefore to consider also the more complicated case of the chain (\ref{1.01})
with $J_n=J$ and $D_n=D_0-(-1)^nD$.
This choice of a dimerized Dzyaloshinskii-Moriya interaction covers both limiting cases
(i) of the alternating-sign Dzyaloshinskii-Moriya interaction when $D_0=0$
and
(ii) of the constant Dzyaloshinskii-Moriya interaction when $D=0$.
Exploiting the transformation (\ref{3.01})
with $\varphi_m=\arctan\left(\left(D_0-(-1)^mD\right)/J\right)$
we arrive at a chain without the Dzyaloshinskii-Moriya interaction
but only with the dimerized $XX$ exchange interaction
$\tilde{J}_n={\rm{sgn}}(J)\sqrt{J^2+\left(D_0-(-1)^nD\right)^2}$.
To find the relation between the $xx$ dynamic structure factor $S_{xx}(\kappa,\omega)$ of the $XX$ chain
with the dimerized Dzyaloshinskii-Moriya interaction
and the dynamic structure factors $\tilde{S}_{\alpha\beta}(\kappa,\omega)$ of the dimerized $XX$ chain
without the Dzyaloshinskii-Moriya interaction
we proceed as follows.
First,
we note that exploiting (\ref{3.01}) in Eq. (\ref{1.02}) yields
\begin{eqnarray}
\label{3.04}
S_{xx}(\kappa,\omega)
=\frac{1}{N} \sum_{j=1}^N\sum_{m=1}^N\exp\left(-{\rm{i}}\kappa m\right)
\int_{-\infty}^{\infty}{\rm{d}}t\exp\left({\rm{i}}\omega t\right)
\nonumber\\
\times
\left(
\cos\left(\phi_{j+m}-\phi_j\right)
\langle \tilde{s}_j^x(t)\tilde{s}_{j+m}^x\rangle
-\sin\left(\phi_{j+m}-\phi_j\right)
\langle \tilde{s}_j^x(t)\tilde{s}_{j+m}^y\rangle
\right).
\end{eqnarray}
After introducing the notations
$\varphi_{{\rm{o}}}=\arctan\left((D_0+D)/J\right)$,
$\varphi_{{\rm{e}}}=\arctan\left((D_0-D)/J\right)$
and
$\varphi^{\pm}=(\varphi_{{\rm{o}}}\pm\varphi_{{\rm{e}}})/2$
we can write
$\phi_{j+m}-\phi_j
=m\varphi^++(-1)^j\left(((-1)^m-1)/2\right)\varphi^-$.
Then after inserting this result into Eq. (\ref{3.04}) and some manipulations
Eq. (\ref{3.04}) becomes
\begin{eqnarray}
\label{3.05}
S_{xx}(\kappa,\omega)
=
\frac{1}{2}\cos^2\frac{\varphi^-}{2}
\left(
\tilde{S}_{xx}(\kappa-\varphi^+,\omega)
+\tilde{S}_{xx}(\kappa+\varphi^+,\omega)
\right.
\nonumber\\
\left.
+{\rm{i}}\tilde{S}_{xy}(\kappa-\varphi^+,\omega)
-{\rm{i}}\tilde{S}_{xy}(\kappa+\varphi^+,\omega)
\right)
\nonumber\\
+
\frac{1}{2}\sin^2\frac{\varphi^-}{2}
\left(
\tilde{S}_{xx}(\kappa\mp\pi-\varphi^+,\omega)
+\tilde{S}_{xx}(\kappa\mp\pi+\varphi^+,\omega)
\right.
\nonumber\\
\left.
+{\rm{i}}\tilde{S}_{xy}(\kappa\mp\pi-\varphi^+,\omega)
-{\rm{i}}\tilde{S}_{xy}(\kappa\mp\pi+\varphi^+,\omega)
\right).
\end{eqnarray}
Eq. (\ref{3.05}) in the limit $D_0=0$ transforms into (\ref{3.03}) since
$\varphi^+=0$,
$\varphi^-=\varphi=\arctan(D/J)$.
Eq. (\ref{3.05}) also contains the result of Ref. \onlinecite{16}
in the limit $D=0$ since
$\varphi^+=\arctan(D_0/J)$,
$\varphi^-=0$.

\begin{figure}
\epsfig{file = 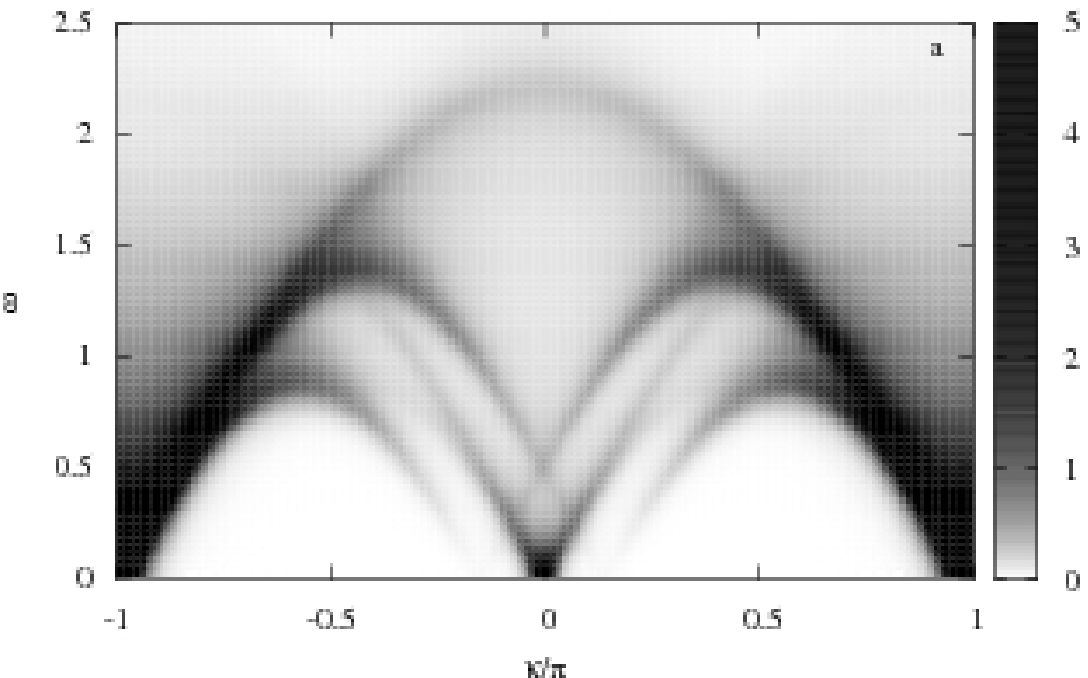, height = 0.25\linewidth}\\
\epsfig{file = 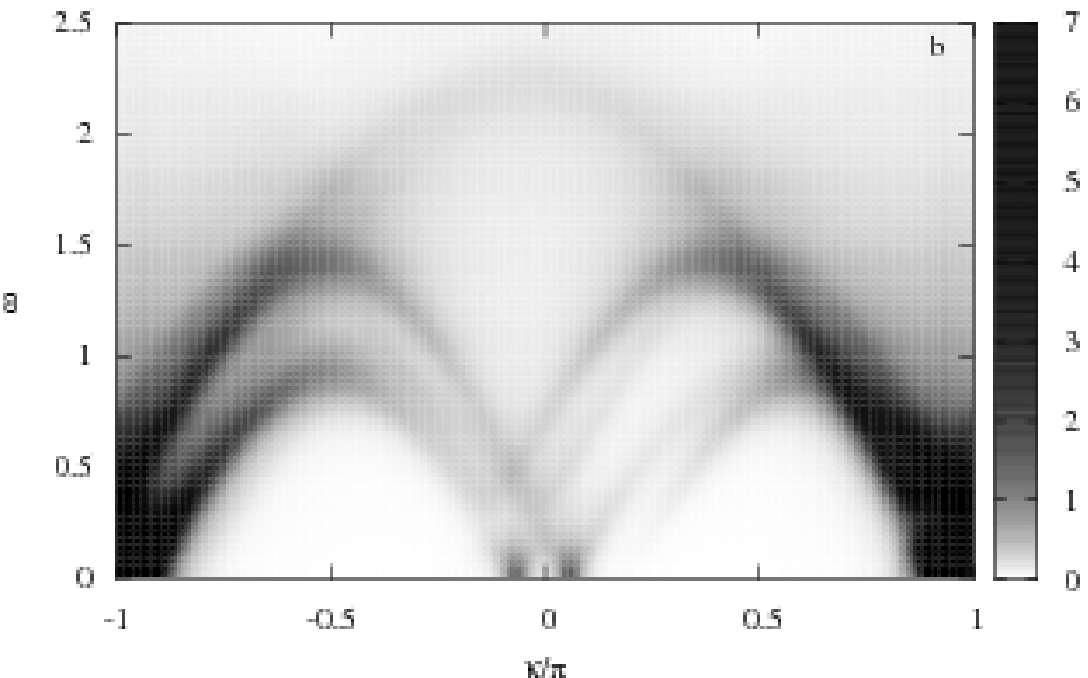, height = 0.25\linewidth}
\caption{$S_{xx}(\kappa,\omega)$
for the chain (\ref{1.01})
with $J_n=1$,
$D_n=D_0-(-1)^nD$,
$D_0=0$ (upper panel a),
$D_0=0.25$ (lower panel b),
$D=0.5$,
$\Omega=0.25$ at low temperature, $\beta=20$.}
\label{fig3}
\end{figure}

In Fig.~\ref{fig3}
we illustrate the effect of the dimerized Dzyaloshinskii-Moriya interaction
on the $xx$ dynamic structure factor at low temperatures.
The panel a corresponds to the case $D_0=0$
($S_{xx}(\kappa,\omega)$ is obtained using Eq. (\ref{3.03})),
whereas the panel b corresponds to the case $D_0\ne 0$
($S_{xx}(\kappa,\omega)$ is obtained using the more general Eq. (\ref{3.05})).

\subsection{Random case}

\begin{figure}
\epsfig{file = 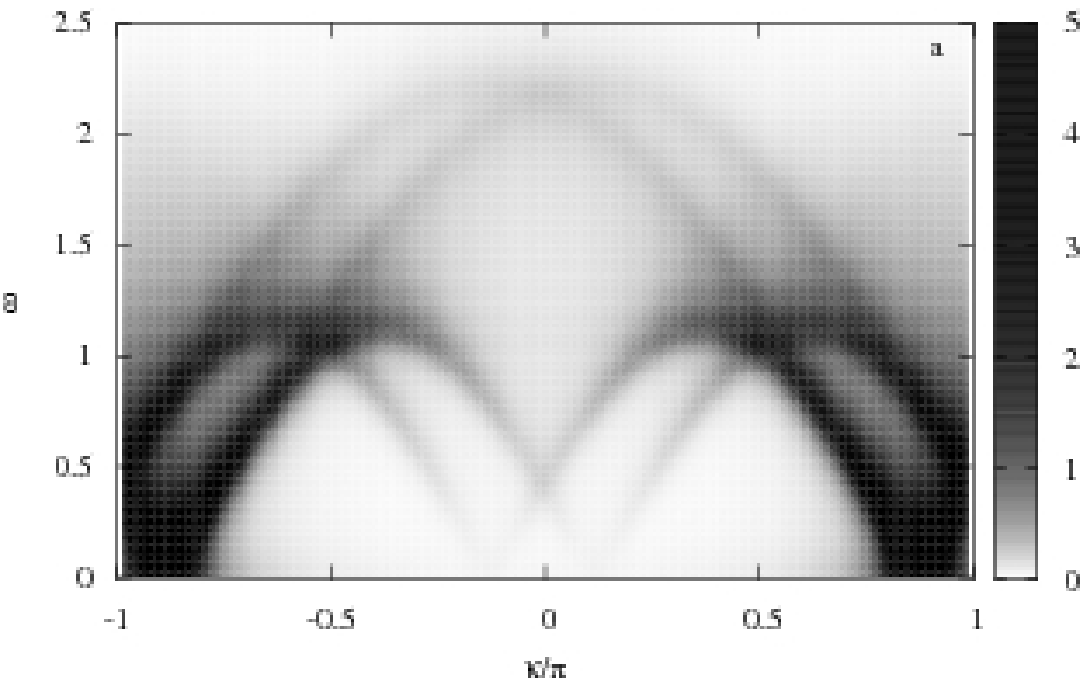, height = 0.25\linewidth}
\epsfig{file = 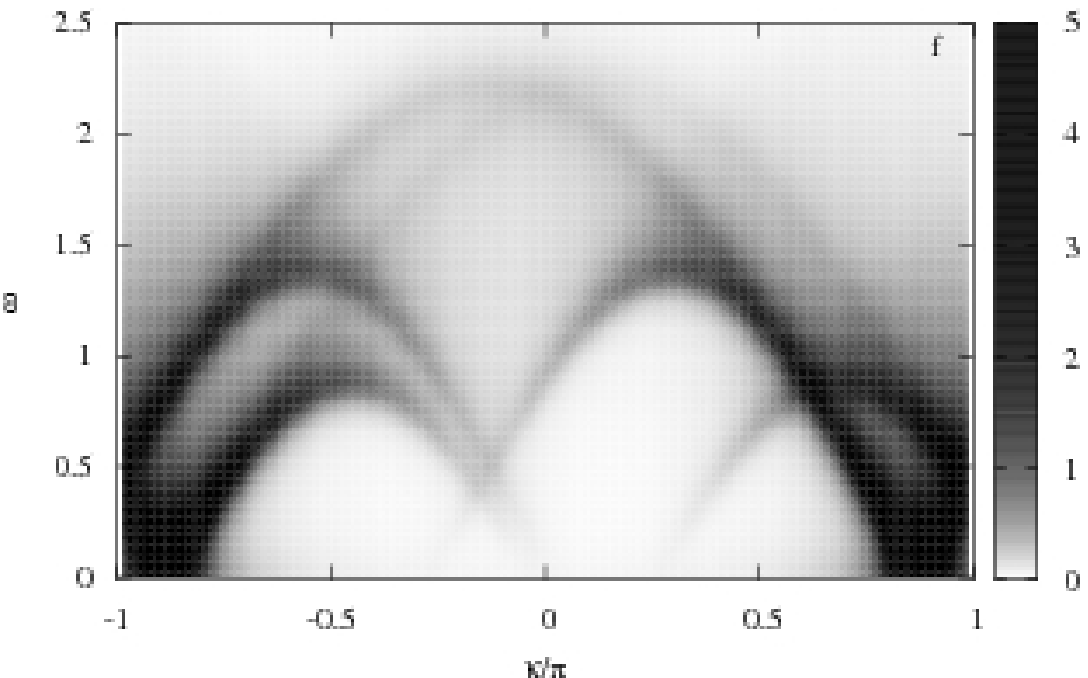, height = 0.25\linewidth}\\
\epsfig{file = 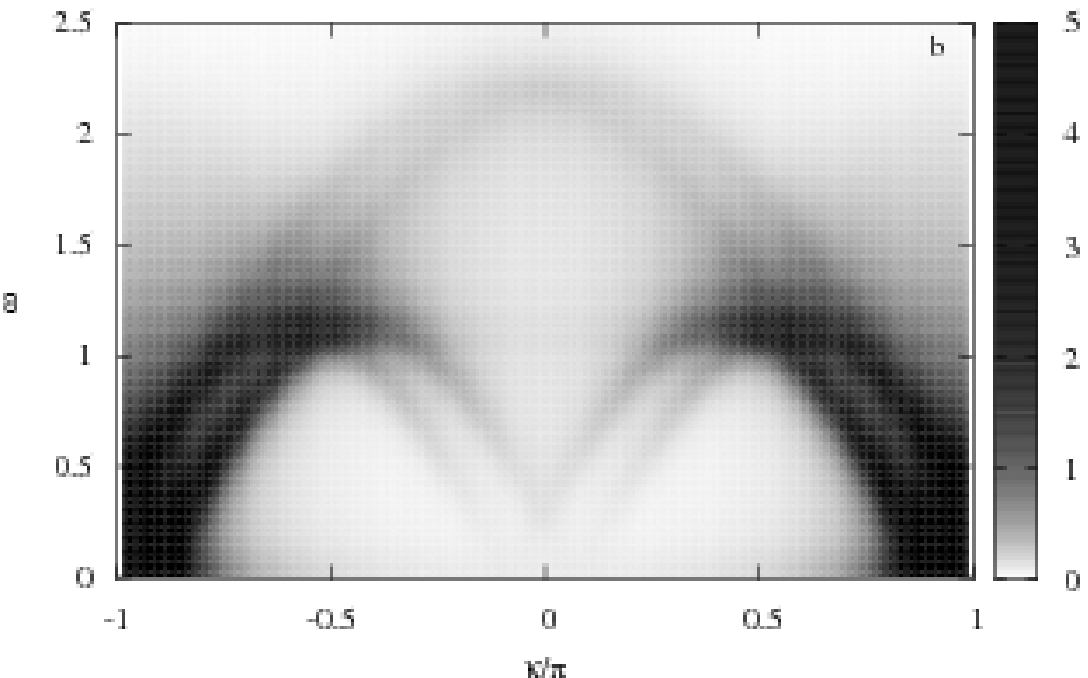, height = 0.25\linewidth}
\epsfig{file = 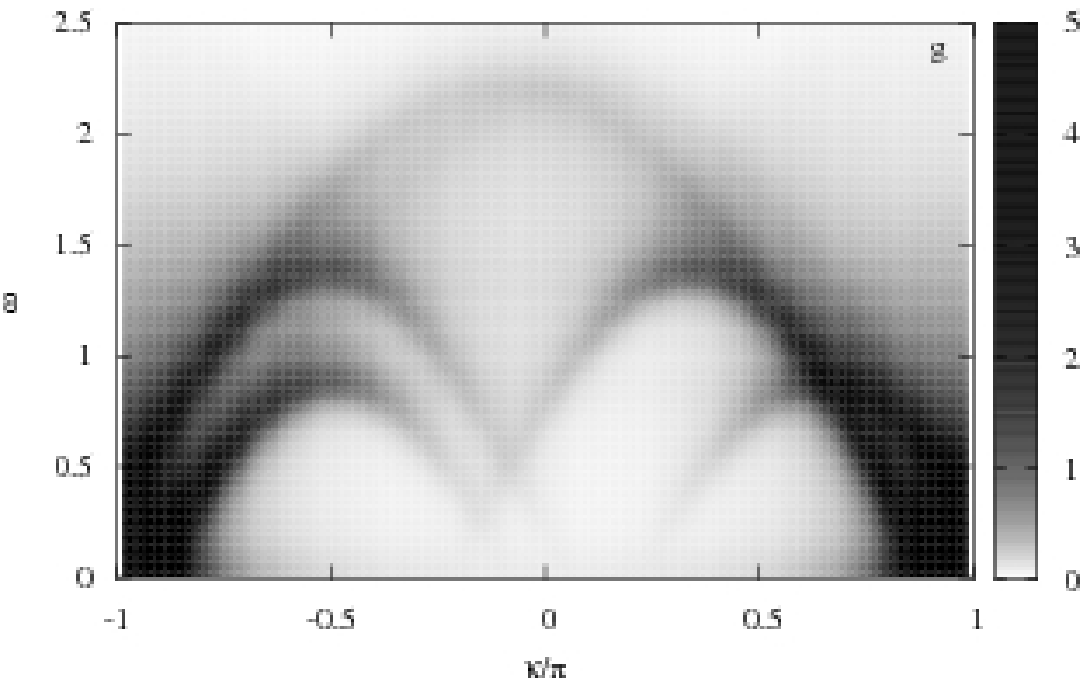, height = 0.25\linewidth}\\
\epsfig{file = 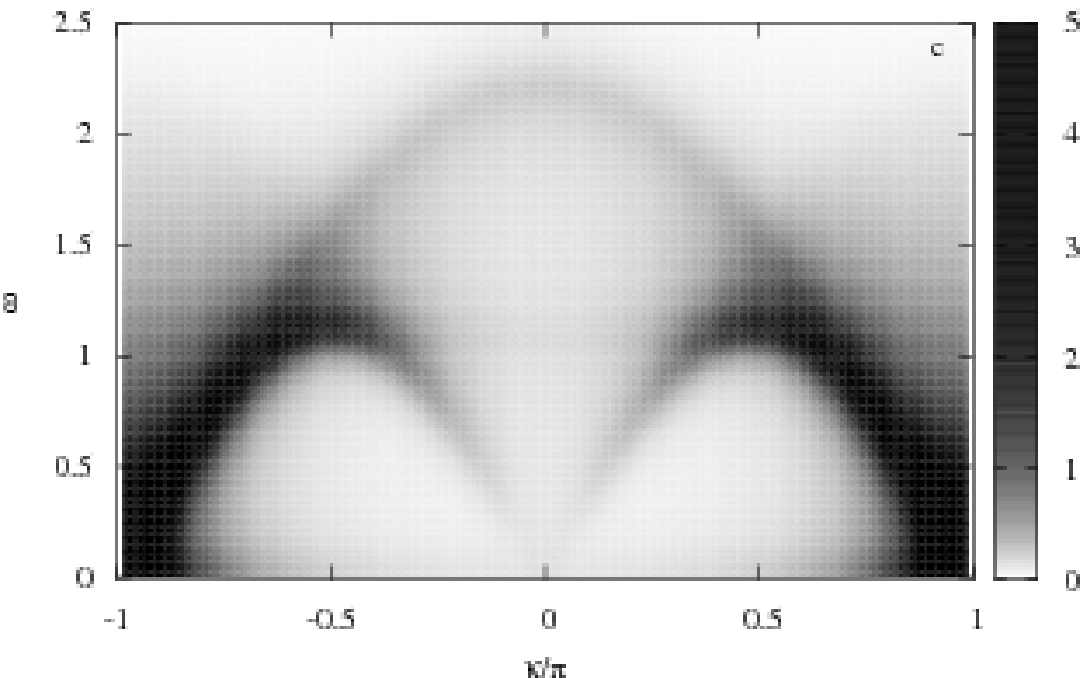, height = 0.25\linewidth}
\epsfig{file = 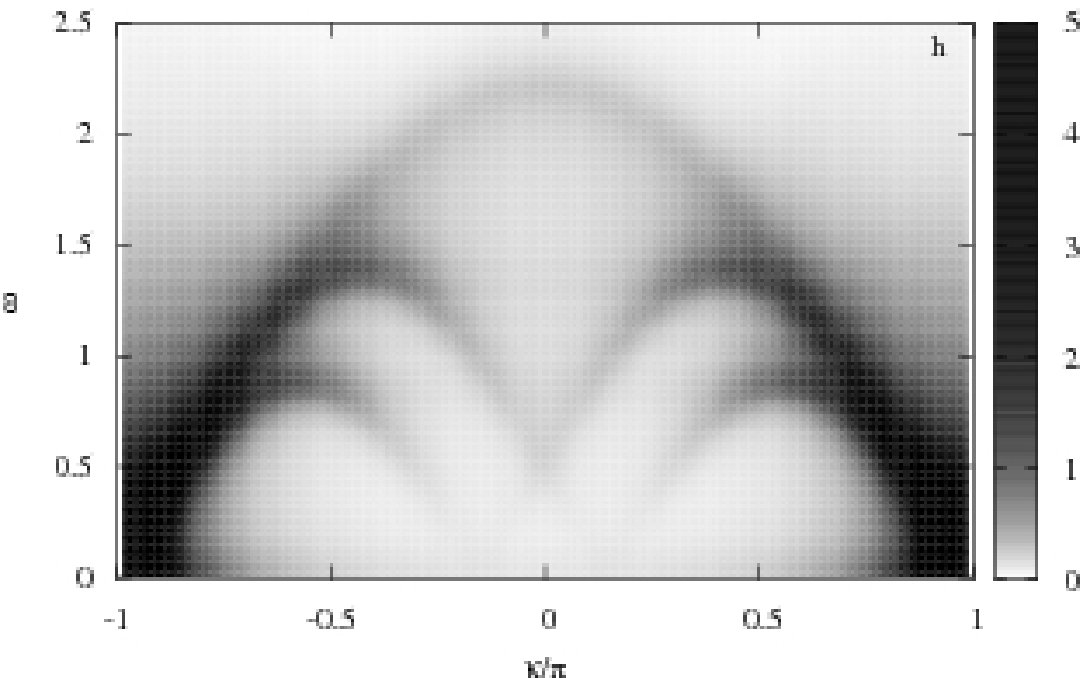, height = 0.25\linewidth}\\
\epsfig{file = 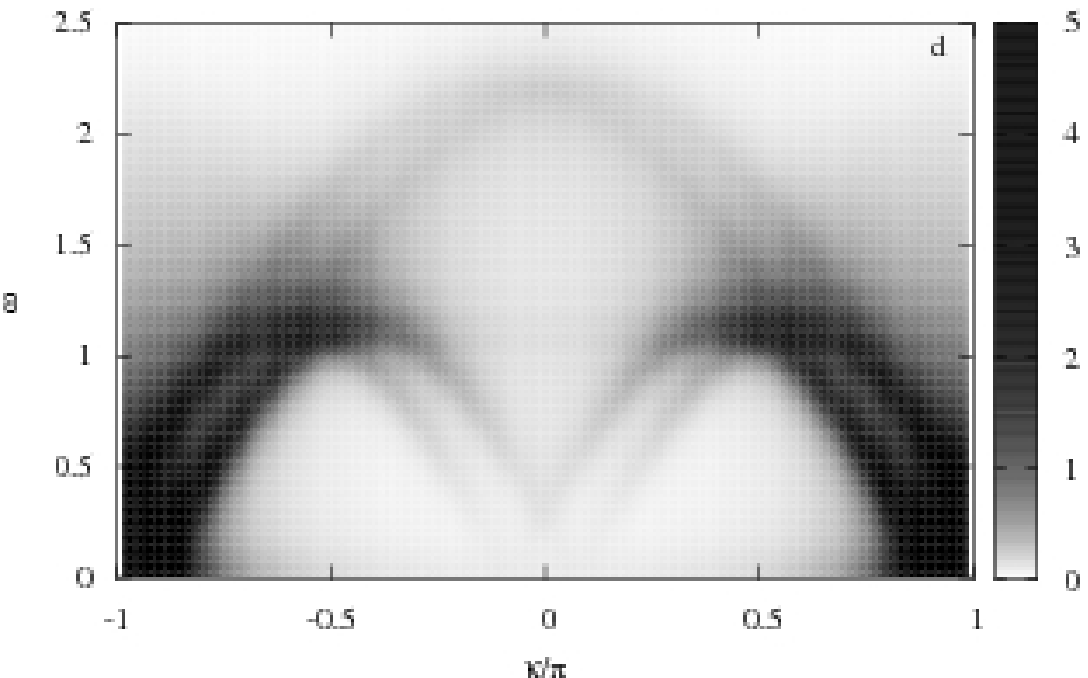, height = 0.25\linewidth}
\epsfig{file = 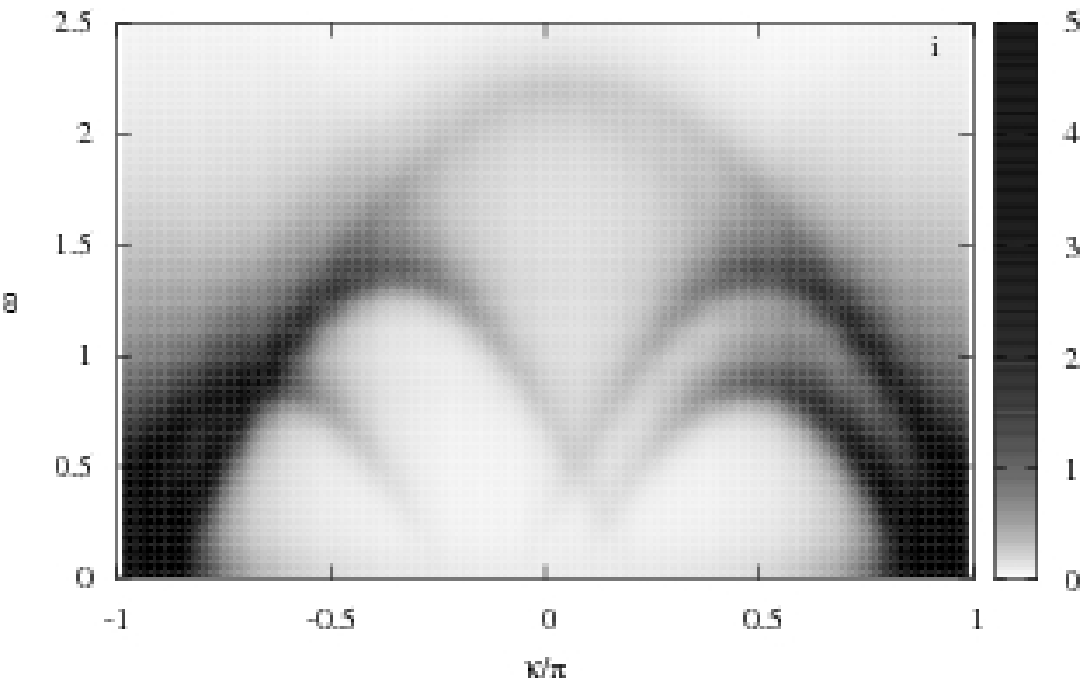, height = 0.25\linewidth}\\
\epsfig{file = 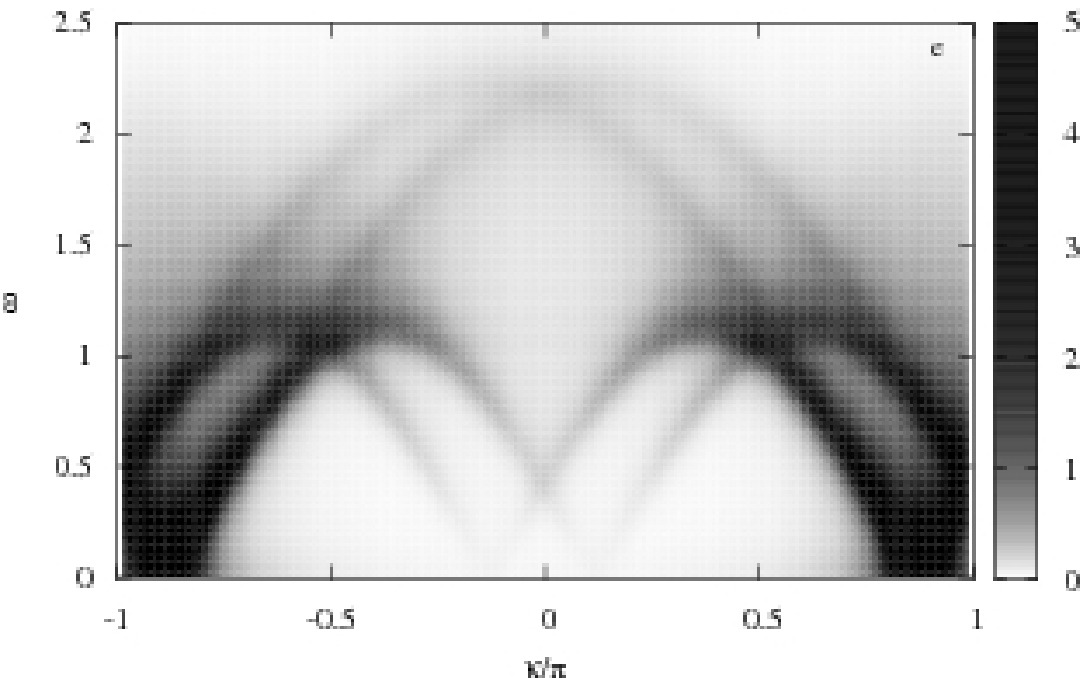, height = 0.25\linewidth}
\epsfig{file = 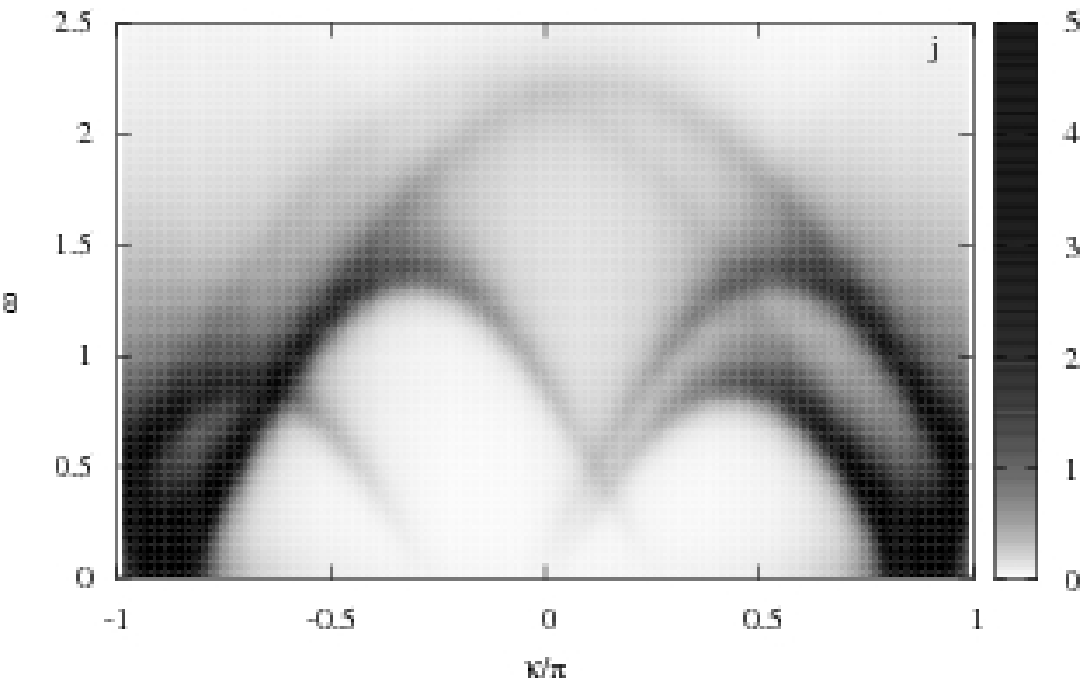, height = 0.25\linewidth}\\
\caption{$\overline{S_{xx}(\kappa,\omega)}$
for the chain (\ref{1.01})
with $J_n=1$,
$D_n=\lambda_nD$,
where $\lambda_n$ is a random variable with the probability distribution (\ref{2.05}).
$D=0.5$,
$\Omega=0.01$ (left panels, a \ldots e),
$\Omega=0.25$ (right panels, f \ldots j),
$\beta=20$.
The values of $p$ in Eq. (\ref{2.05}) are as follows:
from top to bottom
$p=0.1$ (${\xi_D}\approx 26.77$),
$p=0.25$ (${\xi_D}\approx 12.31$),
$p=0.5$ (${\xi_D}\approx 8.96$),
$p=0.75$ (${\xi_D}\approx 12.31$),
$p=0.9$ (${\xi_D}\approx 26.77$).}
\label{fig4}
\end{figure}

Finally,
we pass to the case when the Dzyaloshinskii-Moriya interaction $D_n=\lambda_n D$
is given by a sequence of independent random variables $\{\lambda_n\}$
each with the bimodal probability distribution (\ref{2.05}).
For a specific realization of the signs of the Dzyaloshinskii-Moriya interaction
we can eliminate $D_n$ from the Hamiltonian $H$ (\ref{1.01}) by the transformation (\ref{3.01})
with $\varphi_m=\lambda_m\varphi$,
$\varphi=\arctan(D/J)$ arriving at the model $\tilde{H}$
with only $XX$ exchange interaction $\tilde{J}={\rm{sgn}}(J)\sqrt{J^2+D^2}$.
To calculate the random-averaged $xx$  dynamic structure factor we need
\begin{eqnarray}
\label{3.06}
\overline{\langle s_j^x(t)s_{j+m}^x\rangle}
=
\overline{\cos\left(\phi_{j+m}-\phi_{j}\right)}
\langle \tilde{s}_j^x(t)\tilde{s}_{j+m}^x\rangle
-
\overline{\sin\left(\phi_{j+m}-\phi_{j}\right)}
\langle \tilde{s}_j^x(t)\tilde{s}_{j+m}^y\rangle.
\end{eqnarray}
Noting that
\begin{eqnarray}
\label{3.07}
\overline{\cos\left(\left(\lambda_1+\ldots+\lambda_m\right)\varphi\right)}
=\frac{1}{2}\left(p\exp\left(-{\rm{i}}\varphi\right)+(1-p)\exp\left({\rm{i}}\varphi\right)\right)^m
\nonumber\\
+\frac{1}{2}\left(p\exp\left({\rm{i}}\varphi\right)+(1-p)\exp\left(-{\rm{i}}\varphi\right)\right)^m
\nonumber\\
=
\left(\cos^2\varphi+(1-2p)^2\sin^2\varphi\right)^\frac{m}{2}
\cos\left(m\arctan\left((1-2p)\tan\varphi\right)\right),
\nonumber\\
\overline{\sin\left(\left(\lambda_1+\ldots+\lambda_m\right)\varphi\right)}
=
\left(\cos^2\varphi+(1-2p)^2\sin^2\varphi\right)^\frac{m}{2}
\sin\left(m\arctan\left((1-2p)\tan\varphi\right)\right)
\end{eqnarray}
and introducing the notations
${\xi_D}=-1/\ln\sqrt{\cos^2\varphi+(1-2p)^2\sin^2\varphi}$,
${\varphi_D}=\arctan\left((1-2p)\tan\varphi\right)$
one finds that
\begin{eqnarray}
\label{3.08}
\overline{\langle s_j^x(t)s_{j+m}^x\rangle}
=\exp\left(-\frac{\vert m\vert}{{\xi_D}}\right)
\left(
\cos\left(m{\varphi_D}\right)\langle \tilde{s}_j^x(t)\tilde{s}_{j+m}^x\rangle
-
\sin\left(m{\varphi_D}\right)\langle \tilde{s}_j^x(t)\tilde{s}_{j+m}^y\rangle
\right).
\end{eqnarray}
Using Eq. (\ref{3.08}) the random-averaged $xx$ dynamic structure factor can be written as follows
\begin{eqnarray}
\label{3.09}
\overline{S_{xx}(\kappa,\omega)}
=
\sum_{m=0,\pm1,\pm2,\ldots}
\exp\left(-{\rm{i}}\kappa m-\frac{\vert m\vert}{{\xi_D}}\right)
\int_{-\infty}^{\infty}{\rm{d}}t\exp\left({\rm{i}}\omega t\right)
\nonumber\\
\times
\left(
\cos\left(m{\varphi_D}\right)\langle \tilde{s}_j^x(t)\tilde{s}_{j+m}^x\rangle
-
\sin\left(m{\varphi_D}\right)\langle \tilde{s}_j^x(t)\tilde{s}_{j+m}^y\rangle
\right).
\end{eqnarray}
On the r.h.s. in Eq. (\ref{3.09})
we have the correlation functions of the uniform $XX$ chain with the exchange constant
${\rm{sgn}}(J)\sqrt{J^2+D^2}$.
We use Eq. (\ref{3.09}) to calculate $\overline{S_{xx}(\kappa,\omega)}$
for the model with the random-sign Dzyaloshinskii-Moriya interaction
through the known results for $\langle \tilde{s}_j^{\alpha}\tilde{s}_{j+m}^{\beta}\rangle$ \cite{22,12}.
The results are shown in Fig.~\ref{fig4}.
The correlation length ${\xi_D}$ attains its minimal (nonzero) value
$-1/\ln\vert\cos\varphi\vert$ at $p=1/2$.
A comparison of
Figs.~\ref{fig4}c and \ref{fig4}h
to Figs.~\ref{fig2}a, \ref{fig2}f
shows that $\overline{S_{xx}(\kappa,\omega)}$ looks very similar for weak disorder ($p=0.1$)
in the sign of the exchange interaction on one hand and for maximum disorder ($p=1/2$)
in the sign of the Dzyaloshinskii-Moriya interaction on the other hand.
It looks as if the random-sign Dzyaloshinskii-Moriya interaction at $p=1/2$ does not manifest itself
in $\overline{S_{xx}(\kappa,\omega)}$. This similarity becomes evident if we notice that
the correlation lengths $\xi$ and $\xi_D$ are of the same order
for the considered conditions (see the captions to Figs.~\ref{fig2}, \ref{fig4})
and $\varphi_D$ tends to zero which cancels any signals of the Dzyaloshinskii-Moriya interaction
due to formula (\ref{3.09}).

We can proceed with analytical calculations
for the case $T=0$, $\vert\Omega\vert>\sqrt{J^2+D^2}$.
Comparing Eq. (\ref{3.09}) and Eq. (\ref{2.10})
we see that the result we are interested in follows from Eq. (\ref{2.10})
after the changes
$J\to\tilde{J}={\rm{sgn}}(J)\sqrt{J^2+D^2}$,
$\kappa\to\kappa-{\rm{sgn}}(\Omega){\varphi_D}$.
In particular,
for the nonrandom case when $p=0$ or $p=1$
we recover the result reported in Ref. \onlinecite{16},
$\overline{S_{xx}(\kappa,\omega)}
=(\pi/2)\delta\left(\omega-\vert\Omega\vert-\tilde{J}\cos(\kappa-{\rm{sgn}}(\Omega)\varphi)\right)$.

Interestingly, we can extend the scheme explained above to more
complicated random chains assuming
$J_n={\cal{J}}\cos\left(f(\lambda_n)\right)$,
$D_n={\cal{J}}\sin\left(f(\lambda_n)\right)$, where $f(x)$ is an
arbitrary function, for example, $f(x)=A+Bx$, and $\lambda_n$ is a
random variable with an arbitrary probability distribution
$p(\lambda_n)$ (not necessarily with the bimodal probability
distribution (\ref{2.05})).
After exploiting the transformation (\ref{3.01})
with $\varphi_m=A+B\lambda_m$
we arrive at the Hamiltonian $\tilde{H}$
given by Eq. (\ref{1.01})
with $J_n={\cal{J}}$, $D_n=0$.
$\overline{\langle s_j^x(t)s_{j+m}^x\rangle}$
is again given by Eq. (\ref{3.06}),
however, Eq. (\ref{3.07}) now reads
\begin{eqnarray}
\label{3.10}
\overline{\cos\left(mA+B\left(\lambda_1+\ldots+\lambda_m\right)\right)}
=\vert F(B)\vert^m
\cos\left(m\left(A+\arg(F(B))\right)\right),
\nonumber\\
\overline{\sin\left(mA+B\left(\lambda_1+\ldots+\lambda_m\right)\right)}
=\vert F(B)\vert^m
\sin\left(m\left(A+\arg(F(B))\right)\right),
\end{eqnarray}
where
\begin{eqnarray}
\label{3.11}
F(B)=\int{\rm{d}}\lambda_np(\lambda_n)\exp\left({\rm{i}}B\lambda_n\right)
=\vert F(B)\vert
\exp\left({\rm{i}}\arg(F(B))\right)
\end{eqnarray}
is the characteristic function of the random variable $\lambda_n$.
Now we introduce the notations
${\xi_D}=-1/\ln\vert F(B)\vert$,
${\varphi_D}=A+\arg(F(B))$ and arrive at Eq. (\ref{3.08})
and Eq. (\ref{3.09}). For the model with the
bimodal distribution considered earlier we have  to put
${\cal{J}}={\rm{sgn}}(J)\sqrt{J^2+D^2}$, $A=0$,
$B=\arctan(D/J)=\varphi$, and therefore
$F(B)=p\exp\left(-{\rm{i}}\varphi\right)+(1-p)\exp\left({\rm{i}}\varphi\right)$,
$\vert F(B)\vert = \sqrt{\cos^2\varphi+\left(1-2p\right)^2\sin^2\varphi}$,
$\arg(F(B))=\arctan\left((1-2p)\tan\varphi\right)$ and we
reproduce Eqs. (\ref{3.08}), (\ref{3.09}) with the
expressions for ${\xi_D}$ and ${\varphi_D}$ given just before Eq. (\ref{3.08}).

\section{Conclusions}
\label{sec4}

To summarize,
we have considered a number of inhomogeneous (periodic or random)
spin-1/2 $XX$ chains,
to examine their dynamic properties.
The models considered are distinguished
by the possibility to eliminate the inhomogeneity
from the spin Hamiltonian by a suitable unitary transformation
(see Eqs. (\ref{2.01}), (\ref{3.01}))
and therefore to
reduce the problem to the well known one for the uniform model.
We use exact analytical and precise numerical data to analyze the dynamic structure factors
of the periodic/random spin-1/2 $XX$ chains.
The models considered show rather complex behavior
which, however, can be explained by the corresponding properties of the basic uniform model.
Thus, for the periodic chains
only the correspondingly modified characteristic curves
Eqs. (\ref{1.04}), (\ref{1.06}) are seen
in the complex pattern displayed by the dynamic structure factor at low temperatures.
In the high-temperature limit only Eq. (\ref{1.07}) is relevant.
In the cases considered
the observed complexity has a simple origin.
We also stress here that we have reported
rigorous analytical results for dynamic structure factors of some periodic/random quantum spin chains.
In comparison,
direct numerical treatment of random quantum spin chains
would imply many calculations of dynamic quantities
for different realizations of the random couplings
and a subsequent average over these realizations,
which altogether would require an enormous amount of computer time.

It is interesting to note
that the effects of temperature and of random couplings
on the $xx$/$xy$ dynamic structure factors are different
(compare Eq. (\ref{1.07}) and Eqs. (\ref{2.08}), (\ref{2.11})).
Although in both cases only the autocorrelation function determines the dynamic structure factor
(for sufficiently high temperature or sufficiently strong randomness),
at high temperatures the dynamic structure factor is $\kappa$-independent and shows Gaussian ridges
(see Eq. (\ref{1.07})).
That is due to the Gaussian time decay of the autocorrelation function \cite{09,10}
which should be contrasted
to the slow long-time decay of the autocorrelation function at low temperatures.

The spin chain models discussed in our study are obviously of a rather special kind,
and it would be highly desirable to obtain reliable results
also for more general types of inhomogeneity in the interspin couplings,
where not only the signs but also the absolute values of the couplings vary.
For those more general models,
however,
the present methods are not applicable,
and different methods or approximations have to be employed,
such as in Refs. \onlinecite{30,23}, for example.
The special models treated in our present study will then be useful
in providing a testing ground for the more general (but possibly less reliable) methods
capable of dealing with a broader class of systems.

Finally,
the dynamic structure factors provide
benchmarks for determining interspin interactions.
In our paper we have demonstrated by some examples
how periodic modulations or random variations in the signs of nearest-neighbor interactions
manifest themselves in the dynamic structure factor.
We note that the techniques
used here for $XX$ chains may also be applied
to study dynamic structure factors of more general $XXZ$ chains
with periodic or random sign changes  in the $XY$ part of the interactions,
provided that sufficiently precise data for
the corresponding uniform systems become available.

\section*{Acknowledgments}

This research was supported by  a NATO collaborative linkage grant
(reference number CBP.NUKR.CLG 982540,
project ``Dynamic Probes of Low-Dimensional Quantum Magnets'').
T.~V. and T.~K. acknowledge the kind hospitality of the University of Dortmund
where part of this work was done.

\end{document}